\documentclass[pdflatex,sn-mathphys-num]{sn-jnl}

\usepackage{graphicx}
\usepackage{multirow}
\usepackage{amsmath,amssymb,amsfonts}
\usepackage{amsthm}
\usepackage{mathrsfs}
\usepackage[title]{appendix}
\usepackage{xcolor}
\usepackage{textcomp}
\usepackage{manyfoot}
\usepackage{booktabs}
\usepackage{array}
\usepackage{longtable}
\usepackage[shortlabels]{enumitem}
\usepackage{url}
\theoremstyle{thmstyleone}

\theoremstyle{thmstyletwo}

\theoremstyle{thmstylethree}

\hypersetup{%
  pdftitle={Terrain-Surface-Area Cover from LiDAR: Inverse-Probability-Weighted Estimation and a Global Cram\'er--von Mises Lack-of-Fit Test},
  pdfauthor={Marco Mandap},
  pdfkeywords={terrain-referenced canopy cover, surface-area measure, inverse-probability weighting, Cramer--von Mises, spatial goodness-of-fit, LiDAR detection, formal verification}
}

\begin{document}

\title[Terrain-Surface-Area Cover from LiDAR]{Terrain-Surface-Area Cover from LiDAR: Inverse-Probability-Weighted Estimation and a Global Cram\'er--von Mises Lack-of-Fit Test}

\author*[1]{\fnm{Marco} \sur{Mandap}}\email{marco.mandap@bulsu.edu.ph}

\affil*[1]{\orgdiv{Mathematics Department, College of Science}, \orgname{Bulacan State University}, \orgaddress{\city{City of Malolos}, \state{Bulacan}, \postcode{3000}, \country{Philippines}}}

\abstract{Canopy cover is normally reported as a fraction of a horizontal map plane. When the domain of
inference is the land surface, the appropriate target is the fraction of terrain surface area
under canopy. We define this estimand on a $C^1$ graph surface and show that it differs from
vertical-projection cover by the planimetric covariance of cover with the area density, divided
by the mean area density; the rumple index is a different quantity. A LiDAR pulse is recorded as
a canopy return only when detection succeeds, so the naive return fraction estimates cover
multiplied by the canopy detection probability. We study an inverse-probability-weighted
estimator that uses that probability as the inclusion probability, and a weighted
corner-averaged Cramér–von Mises lack-of-fit statistic for constant mean surface-area cover
under independent canopy returns, calibrated by a parametric bootstrap. Synthetic validation
on the flat plane and the Gaussian hill under the phenomenological detection model gives a naive
bias near $-0.40$ and a weighted bias near $0.00$; at 5000 pulses the test holds size at $0.02$–$0.06$ across the simulated null
cells and reaches power $0.79$ under surface-uniform and $0.85$ under plane-uniform sampling
against broad slope-dependent cover, but only $0.09$–$0.13$ against a compact hotspot.
The surface-uniform definitions and their exact reductions are machine-checked in Lean 4. No field data yet.
}


\maketitle

\section{Introduction}
\label{sec:intro}

Canopy cover is usually reported as a fraction of a horizontal map plane. The standard definition is the proportion of the ground surface covered by the vertical projection of canopy material. The definitional distinction among canopy cover, canopy closure and related measures is set out by \citet{jennings1999canopy}, and \citet{korhonen2011airborne} carry it to discrete-return LiDAR; cover is not the same as angular canopy closure or leaf area index. A definition that fixes the projection also fixes the denominator: map area. Operational products inherit both choices; \citet{li2024vertical} validate the GEDI vertical canopy cover profile and report that total cover correlates with reference measurements but is systematically underestimated. Whether map area is the right denominator is a separate modelling decision, not a measurement detail.

That decision matters when the inferential domain is the land surface. \citet{panek2026novel} argue that two-dimensional projections bias area-based assessments in forested mountain terrain, and construct a three-dimensional framework for effective vegetated surface area from multi-source elevation and canopy height data. Their target is surface area, not a cover fraction, and they do not test the constancy of cover; the estimand developed here is adjacent to their framework rather than contained in it. The same premise, that the process of interest acts on the terrain, motivates terrain-referenced denominators in environmental applications where the relevant area is that of the land surface.

LiDAR makes the denominator problem harder, because the return process need not sample either the plane or the surface uniformly. Scan geometry, pulse loss and occlusion are first-order effects. \citet{ma2017comparison} report that uncertainty in cover estimation increases sharply once scan angles exceed about $12^\circ$, and \citet{dayal2022investigation} investigate scan-angle impacts on stand-attribute predictions and evaluate voxelisation to normalise them. \citet{liu2019improving} show that non-returning pulses bias cover estimation and that a loss-ratio correction improves accuracy. \citet{xu2026quantifying} show that canopy clumping changes detectability at fixed canopy density. Terrain enters twice: slope and off-nadir pointing distort waveform returns \citep{park2014application}, and undulating terrain changes point density, beam footprint and off-nadir scan angle, so that leaf-area-index discrepancies intensify with steepness \citep{li2023topography}. Slope-adaptive ground-return corrections have been proposed for GEDI cover retrieval \citep{hou2026slope}, and occlusion is structural rather than incidental in dense stands \citep{schneider2019quantifying,donager2021adjudicating}. Detection is therefore not a nuisance to be averaged away; it is a function of the terrain and canopy that the cover fraction is meant to describe.

The statistical machinery is established as well. The Cram\'er--von Mises criterion is classical; it admits weighted forms and forms with estimated parameters \citep{ozturk1997generalised,martynov2011weighted}, and goodness of fit on curved spaces is developed: \citet{jupp2005sobolev} gives coordinate-invariant Sobolev tests on compact Riemannian manifolds, \citet{jupp2020measures} extend the construction through almost-canonical transformations, and \citet{garcia2021cvm} give a projection-based Cram\'er--von Mises test of uniformity on the hypersphere. For spatial patterns, \citet{zimmerman1993bivariate} introduced the corner-averaged bivariate Cram\'er--von Mises test for spatial randomness, and Cram\'er--von Mises-type tests now address separability of marks and locations \citep{schoenberg2004testing} and mark independence \citep{grabarnik2011correct}, while \citet{myllymaki2015deviation} compare the power of deviation tests for marked patterns. The lesson is double-edged: the functional is available, and the null must match the resampling scheme. A permutation test of exchangeability is not a test of constant mean cover under spatial correlation \citep{grabarnik2011correct}.

The gap is specific. We did not find, in the literature we searched, a study that defines a cover fraction with respect to terrain surface area on a graph surface and tests its constancy with a detection-weighted Cram\'er--von Mises statistic. Goodness of fit for the mark law itself exists, with a parametric mark model \citep{mrkvicka2012goodness} or the Palm mark distribution of a stationary process \citep{heinrich2014asymptotic}, but not with a terrain surface reference measure and detection weights. The search was a structured scan of statistics, remote-sensing and ecology sources, not a systematic review, so the negative claim is bounded by it. The nearest work estimates surface area in mountain terrain \citep{panek2026novel}, estimates cover as a vertical projection \citep{li2024vertical}, or tests random labelling and mark independence without a surface-area reference measure \citep{zimmerman1993bivariate,grabarnik2011correct}. What is missing is the combination: an estimand weighted by terrain surface area, an estimator that corrects for the probability that a canopy return is recorded, and a lack-of-fit statistic whose reference measure is the same surface measure.

The paper makes four contributions, addressing what is estimated, how it can be estimated from returns, and whether its mean is constant. Evidence classes are named in parentheses: \emph{formalized} denotes a Lean~4 declaration, \emph{simulated} a synthetic Monte Carlo result, \emph{literature} a published source, and \emph{derived} algebra proved in the text. The gap statement rests on the literature search; the four contributions are this paper's own.

\begin{enumerate}
\item[(C1)] We define the terrain-surface-area cover fraction on a $C^1$ graph surface and show that it differs from vertical-projection cover by the planimetric covariance of cover with area density divided by mean area density; the identity is derived in \S\ref{sec:estimand}, checked against two controls on the simulated surfaces, and supported by a machine-checked construction of the surface measure (derived, simulated, formalized).

\item[(C2)] We show that the naive canopy-return fraction estimates the detection-weighted mean $\mathbb{E}_\mu[p\,q_c]$ over emitted pulses, where $q_c$ is the canopy detection probability, rather than $p_{\mathrm{surf}}$, and that inverse-probability weighting with $q_c$ removes the first-order detection bias when $q_c$ is known \citep{horvitz1952generalization}; the finite-sample bounds are formalized and the bias correction is simulated (derived, formalized, simulated).

\item[(C3)] We define a global detection-weighted corner-averaged Cram\'er--von Mises statistic for constant mean surface-area cover under independent canopy returns, calibrate it by a parametric bootstrap, and report its size and power on synthetic terrain; the surface-uniform statistic and its flat, no-detection and all-canopy reductions are formalized, while the operating range is simulated (formalized, simulated).

\item[(C4)] We give a machine-checked specification of the graph-surface measure, the weighted statistic and the reductions between them in Lean~4 \citep{demoura2021lean4,mathlib2020lean}, so that later asymptotic work cannot silently change the object being tested (formalized).
\end{enumerate}

We do not claim a general non-flat reduction of the surface statistic to the cube statistic, a design-based asymptotic null law, local hotspot detection, or validation on real LiDAR or field data; the proved reductions are flat-surface, no-detection or all-canopy reductions, and the calibration is a parametric bootstrap.

\section{Terrain-surface-area cover}
\label{sec:estimand}

\subsection{The graph surface and its area element}

Let $D=[0,1]^2$ and let $f:D\to\mathbb{R}$ be continuously differentiable. The terrain is the graph $S=\{(u,v,f(u,v)):(u,v)\in D\}$. At a point with partial derivatives $f_u$ and $f_v$, the surface-area element is
\begin{equation}
dA(u,v)=\sqrt{1+f_u(u,v)^2+f_v(u,v)^2}\,du\,dv=:\rho(u,v)\,du\,dv.
\label{eq:area-element}
\end{equation}
The function $\rho$ is the area density. Since the radicand is at least one, $\rho\ge 1$, with equality only where the gradient vanishes; $\rho$ is continuous and strictly positive on the compact square. The total surface area is $|S|=\int_D \rho\,du\,dv$, which is finite and positive. Summing triangle areas to obtain the surface area of a terrain model is standard geomorphometry \citep{jenness2004calculating}; the step that is not standard is using that area as the denominator of a cover fraction. On the flat graph $\rho\equiv 1$. On the constant slope $f=au+bv$, $\rho=\sqrt{1+a^2+b^2}$ is constant, so every planimetric patch is stretched by the same factor.

\subsection{The surface-area cover fraction}

\begin{figure}[t]
\centering
\includegraphics[width=\textwidth]{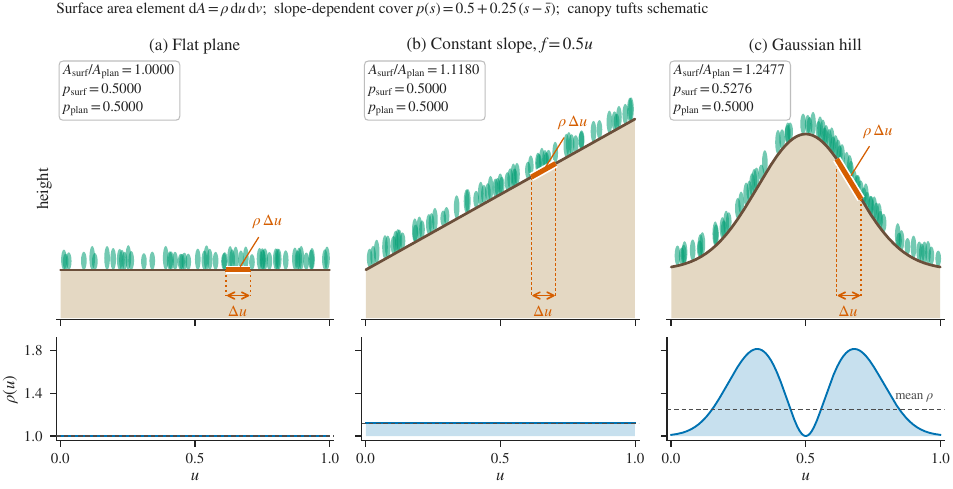}
\caption{The surface-area estimand on the three synthetic surfaces. Top row: terrain
profile at \(v=0.5\) with a schematic canopy (tufts are a drawing device, not a sample),
the surface element \(\rho\,\Delta u\) and its planar shadow \(\Delta u\). Bottom row: the
area density \(\rho(u)\) along the same profile with its mean \(\bar\rho\) (dashed), which
is the area ratio \(A_{\mathrm{surf}}/A_{\mathrm{plan}}\). Cover is the slope-dependent map
\(p(u,v)=0.5+0.25\,(\operatorname{slope}(u,v)-\overline{\operatorname{slope}})\), so it co-varies with area density on the hill and is
constant on the other two surfaces. Consequently \(\bar\rho=1.1180\) on the constant slope
but \(p_{\mathrm{surf}}=p_{\mathrm{planar}}=0.5000\), while the hill has
\(\bar\rho=1.2477\) and \(p_{\mathrm{surf}}=0.5276\) against \(p_{\mathrm{planar}}=0.5000\).
Boxed values are exact grid evaluations, not Monte Carlo estimates.}
\label{fig:estimand}
\end{figure}

Let $C(s)\in\{0,1\}$ be the canopy indicator at surface point $s$, and write $C(u,v)$ for its value at the point of $S$ above $(u,v)$. The normalised surface measure on $D$ is
\[
\mu(A)=\frac{\int_A \rho(u,v)\,du\,dv}{\int_D \rho(u,v)\,du\,dv},
\qquad A\subseteq D .
\]
Because $\rho$ is continuous and strictly positive, $\mu$ is a probability measure with density proportional to $\rho$. The terrain-surface-area cover fraction is
\begin{equation}
p_{\mathrm{surf}}
=\frac{\int_S C(s)\,dA(s)}{\int_S dA(s)}
=\frac{\int_D C(u,v)\,\rho(u,v)\,du\,dv}{\int_D \rho(u,v)\,du\,dv}
=\mathbb{E}_\mu[C].
\label{eq:psurf}
\end{equation}
It is the fraction of the land surface that is under canopy, each terrain patch weighted by its true area. If pulse locations were drawn from $\mu$ and the canopy indicator were observed at each pulse, the sample mean would be unbiased for $p_{\mathrm{surf}}$; the detection problem studied in Section~\ref{sec:sampling} is that the observed canopy-return law is not $\mu$. The geometry is machine-checked: positivity and continuity of $\rho$, finiteness and positivity of $|S|$, the normalisation $\mu(D)=1$, and the reduction of $\mu$ to Lebesgue measure on the flat graph are formalized in Lean~4 (Section~\ref{sec:lean}). Figure~\ref{fig:estimand} shows the three synthetic surfaces and the area density along a profile.

\subsection{What the estimand is not}

The estimand is easily confused with neighbouring quantities. Vertical-projection cover uses planimetric area,
\[
p_{\mathrm{planar}}=\int_D C(u,v)\,du\,dv=\mathbb{E}_{\mathrm{Leb}}[C],
\]
the fraction of the map plane covered by the vertical projection of canopy. It is the quantity used in standard cover definitions and operational products \citep{korhonen2011airborne,li2024vertical}; angular canopy closure and leaf area index are different again \citep{korhonen2011airborne}. Rumple is the ratio of canopy outer surface area to ground area \citep{kane2010comparisons}. It measures the roughness of the canopy outer surface, whereas $p_{\mathrm{surf}}$ measures how much of the terrain surface is covered. The two answer different questions, and this paper reports no rumple values; a quantitative comparison on real terrain is future work. Finally, $p_{\mathrm{surf}}$ is not a rescaling of $p_{\mathrm{planar}}$. It is a ratio of integrals with the same area weight in numerator and denominator, and the next subsection shows that the area ratio alone does not map one to the other.

One caveat is structural. Both the canopy indicator and the terrain point are indexed by the same $(u,v)$, so canopy is still credited vertically: the surface-area denominator changes, not the projection direction. A fully oblique or overhang-aware definition would require a three-dimensional canopy representation and is outside this paper. The estimand is therefore a terrain-area-weighted vertical projection, not a view-angle-dependent canopy closure.

\subsection{When the two fractions differ}

Assume $C$ and $\rho$ are integrable; a bounded indicator $C$ and continuous $\rho$ on the compact square suffice. Write $\mathbb{E}_{\mathrm{planar}}[g]=\int_D g\,du\,dv$ and $\operatorname{Cov}_{\mathrm{planar}}(C,\rho)=\mathbb{E}_{\mathrm{planar}}[C\rho]-\mathbb{E}_{\mathrm{planar}}[C]\,\mathbb{E}_{\mathrm{planar}}[\rho]$. Since $D$ has unit area, $p_{\mathrm{planar}}=\mathbb{E}_{\mathrm{planar}}[C]$. Therefore
\[
p_{\mathrm{surf}}-p_{\mathrm{planar}}
=\frac{\mathbb{E}_{\mathrm{planar}}[C\rho]}{\mathbb{E}_{\mathrm{planar}}[\rho]}-\mathbb{E}_{\mathrm{planar}}[C]
=\frac{\operatorname{Cov}_{\mathrm{planar}}(C,\rho)}{\mathbb{E}_{\mathrm{planar}}[\rho]},
\]
that is,
\begin{equation}
p_{\mathrm{surf}}
=p_{\mathrm{planar}}
+\frac{\operatorname{Cov}_{\mathrm{planar}}(C,\rho)}
{\mathbb{E}_{\mathrm{planar}}[\rho]}.
\label{eq:gap}
\end{equation}
The identity is exact for any integrable $C$ and any positive continuous $\rho$. It uses no independence assumption and no small-slope approximation. The denominator is the mean area ratio, so the quotient is on the same scale as the two fractions and is directly interpretable as a cover difference.

Three consequences follow. First, a non-unit area ratio is not enough to create a gap. If $\rho$ is constant, the covariance vanishes and the two fractions coincide, however steep the terrain. Second, the sign of the gap is the sign of the planimetric covariance of cover with area density. Because $\rho$ increases with the gradient magnitude, cover that rises with slope produces a positive covariance and $p_{\mathrm{surf}}>p_{\mathrm{planar}}$; cover that falls with slope produces a negative one. A negative gap is therefore equally possible, and the identity implies no general inequality between the two fractions. Third, the gap is not a rescaling: $p_{\mathrm{surf}}$ is neither $p_{\mathrm{planar}}\mathbb{E}_{\mathrm{planar}}[\rho]$ nor $p_{\mathrm{planar}}/\mathbb{E}_{\mathrm{planar}}[\rho]$, and knowing $p_{\mathrm{planar}}$ and the area ratio does not determine $p_{\mathrm{surf}}$. The covariance term is the additional functional, and it is the quantity worth reporting as a diagnostic.

\subsection{Two controls}

The identity makes two predictions that require no fitting. On the constant slope $f=0.5u$, the area density is constant at $\rho=\sqrt{1.25}=1.118034$, so the area ratio is $1.1180$, but the slope-dependent cover map $p(u,v)=0.5+0.25(\mathrm{slope}(u,v)-\overline{\mathrm{slope}})$, with $\overline{\mathrm{slope}}$ the grid mean, degenerates to the constant $0.5$, so the covariance is zero and $p_{\mathrm{surf}}=p_{\mathrm{planar}}=0.5000$; the recorded gap is $0.0000$. A steep surface with a large area ratio therefore produces no gap when the area density is constant, however cover varies. On the Gaussian hill $f=0.45\exp(-((u-0.5)^2+(v-0.5)^2)/(2\cdot 0.18^2))$, the same cover map is genuinely slope-dependent, and the covariance is positive. The planimetric covariance of cover with area density is $0.03448$ and the mean area density is $1.24770$; their quotient is $0.027634$, matching the difference recorded in the synthetic study between $p_{\mathrm{surf}}=0.527634$ and $p_{\mathrm{planar}}=0.500000$. The division is an algebraic identity, not an independent estimate; the independent check is the constant-slope control above, where the ratio is $1.1180$ and the gap is zero. The covariance and the mean density are exact grid evaluations on a $500\times 500$ lattice, not Monte Carlo estimates.

\subsection{Magnitude and interpretation}

The difference is modest: $52.76\%$ against $50.00\%$, or $2.76$ percentage points. The hill is a gentle test case---maximum height $0.45$ over the unit square, Gaussian scale $\sigma=0.18$, area ratio $1.2477$---although its flanks are locally steep; the steep annulus is narrow, so the area ratio is moderate. The cover map is $0.5+0.25(\mathrm{slope}-\overline{\mathrm{slope}})$. The value of the identity is not the size of the gap. It is that the gap is an identifiable functional---the covariance of cover with area density, divided by mean area density---that can be reported alongside the two fractions. On steeper or more heterogeneous terrain, or with a stronger association between cover and slope, the identity locates any larger gap in the same covariance; the present study does not estimate that covariance for real terrain.

Section~\ref{sec:sampling} turns to the observation problem: $C$ is not observed directly, only through LiDAR returns whose probability depends on the same terrain and canopy.

\section{Sampling, detection, and estimation}
\label{sec:sampling}

A LiDAR survey observes canopy cover indirectly: it emits pulses at nominal positions on the terrain and records, for each, whether a canopy return was obtained. Two mechanisms intervene between the canopy field and the data---where the pulses fall, and whether a return survives detection. This section specifies both, shows that they require two distinct weights, and isolates the bias of the naive return fraction.

\subsection{The pulse process and the canopy return}
\label{sec:pulse-process}

Let $X_1,\dots,X_n\in D$ be the parameter coordinates of the $n$ emitted pulses and $s_i=(X_i,f(X_i))$ the corresponding surface points. The scan geometry of pulse $i$---incidence angle, range, footprint---is recorded in $\theta_i$ and enters what follows only through the detection probability. For pulse $i$, define the canopy return indicator
\begin{equation}
R_i=C(s_i)\,\mathbf{1}\{\text{a canopy return is recorded for pulse }i\},
\label{eq:canopy-return}
\end{equation}
so that $R_i=1$ exactly when the pulse both strikes canopy and yields a recorded canopy return. Ground-only returns contribute $R_i=0$ and are not used.

Write $p(s)=\mathbb{P}\{C(s)=1\}$ and let $\mu$ be the normalised surface measure of Section~\ref{sec:estimand}, with area density $\rho=(1+f_u^2+f_v^2)^{1/2}$ and total mass $Z=\int_D\rho$. The estimand is the surface-area fraction $p_{\mathrm{surf}}=\int_D p\,d\mu$, which for a realised canopy field is $\int_D C\,d\mu$.

Detection is summarised by $q_c(s,\theta)\in(0,1]$, the probability that a pulse emitted at $s$ under geometry $\theta$ produces a recorded canopy return, given that canopy is present at $s$. When the scan geometry is fixed, we write $q_c(X_i)$ for $q_c(X_i,\theta_i)$.
We assume
\begin{equation}
\mathbb{P}\{R_i=1\mid X_i=s,\,C(s)=c\}=c\,q_c(s,\theta_i).
\label{eq:detection}
\end{equation}
Equation~\eqref{eq:detection} states that detection is conditionally independent of the cover label given location and scan geometry. In the simulations it holds by construction, because the terrain-derived density proxy that drives $q_c$ is independent of the realised binary canopy. In a real acquisition it is an approximation whose quality depends on how much of the detection process geometry and auxiliary structure explain.

\subsection{Two weights}
\label{sec:two-weights}

Let $\xi$ be the design measure of the emitted pulses on $D$, with density $g$ with respect to Lebesgue measure. Two designs matter here. Under \emph{surface-uniform} sampling $g=\rho/Z$, so $\xi=\mu$ and a pulse is as likely to fall on one unit of surface area as on another. Under \emph{plane-uniform} sampling $g\equiv1$, so pulses are uniform on the map square and steep terrain receives fewer pulses per unit of surface area.

One weight covers both. Put
\begin{equation}
w(s)=\frac{\rho(s)/Z}{g(s)},
\label{eq:geometric-weight}
\end{equation}
so that $w\equiv1$ under surface-uniform sampling and $w=\rho/Z$ under plane-uniform sampling. The sampling weight of pulse $i$ is $w(X_i)/q_c(X_i,\theta_i)$, and its two factors answer different questions: $w$ converts a plane-referenced count into a surface-referenced one, while $1/q_c$ converts an emitted pulse into the canopy return it would have produced under complete detection.

The estimator is the Horvitz--Thompson mean over emitted pulses \citep{horvitz1952generalization}. Under surface-uniform sampling, where $w\equiv1$, it is
\begin{equation}
\hat p=\frac{1}{n}\sum_{i=1}^{n}\frac{R_i}{q_c(X_i,\theta_i)},
\label{eq:ipw}
\end{equation}
which is the form studied in the formal core and in the surface-uniform simulations. Under plane-uniform sampling the same estimator carries the geometric weight,
\begin{equation}
\hat p_{\mathrm{plane}}=\frac{1}{n}\sum_{i=1}^{n}\frac{\rho(X_i)}{Z}\,\frac{R_i}{q_c(X_i,\theta_i)} ,
\label{eq:ipw-plane}
\end{equation}
and this is the arm that exercises $\rho$. It is also closer to a real airborne design, in which pulses are not distributed in proportion to surface area; it was run at $n=5000$ (Section~\ref{sec:plane-results}).

The justification is one line of expectation, taken conditional on the canopy field. Since $wg=\rho/Z$ and $\mathbb{E}\{R_i\mid X_i,C\}=C(X_i)\,q_c(X_i,\theta_i)$,
\begin{equation}
\begin{aligned}
\mathbb{E}[\hat p\mid C]
&=\frac{1}{n}\sum_{i=1}^{n}\mathbb{E}_\xi\!\left[\frac{w(X_i)}{q_c(X_i,\theta_i)}\,C(X_i)\,q_c(X_i,\theta_i)\right]\\
&=\frac{1}{n}\sum_{i=1}^{n}\mathbb{E}_\xi\bigl[w(X_i)\,C(X_i)\bigr]
=\int_D C\,d\mu=p_{\mathrm{surf}} .
\end{aligned}
\label{eq:unbiased}
\end{equation}
The inner expectation averages over the pulse location and the detection draw, so the two detection factors cancel and $w$ converts the design measure into surface measure. This is a design-unbiasedness identity conditional on the realised field, not a theorem of the formal core: the roadmap of \texttt{SurfaceCoverIPW.lean} lists unbiasedness and consistency as pending, and Section~\ref{sec:lean} records what is and is not checked. The simulations draw $C$ from a Bernoulli canopy model and compare $\hat p$ with the realised grid value of $p_{\mathrm{surf}}$ for the field drawn; conditional on that field, \eqref{eq:unbiased} is the statement being checked. Design-based estimation, in which the sample carries the inference and models enter as auxiliary information, is the classical frame for large-area forest surveys \citep{stahl2016models}; the estimator here is design-based in the pulse sample and model-based in its weights, closest to their model-assisted frame. Averaging over the canopy model as well, the variance under independent returns is
\begin{equation}
\operatorname{Var}(\hat p)=\frac{1}{n}\left(\mathbb{E}_\xi\!\left[\frac{p(X_i)\,w(X_i)^2}{q_c(X_i,\theta_i)}\right]-p_{\mathrm{surf}}^2\right),
\label{eq:ipw-var}
\end{equation}
Under that independence, for the surface-uniform estimator \eqref{eq:ipw}, where $w\equiv1$, and constant $p$, it exceeds the binomial variance $p(1-p)/n$ whenever $q_c<1$: the weights remove the bias of detection but not its information cost, so the effective sample size lies below $n$. Correlated labels add covariance terms that \eqref{eq:ipw-var} omits: with $U_i=w(X_i)R_i/q_c(X_i,\theta_i)$, the variance carries $\sum_{i\ne j}\operatorname{Cov}(U_i,U_j)$ in addition to the independent-case terms. With them neither \eqref{eq:ipw-var} nor the binomial comparison holds as written. This is why the pulse-density requirement in Section~\ref{sec:validation} concerns information, not $n$.

When the pulse count is not fixed by design, the ratio form
\begin{equation}
\hat p_{\mathrm{ratio}}=\frac{\sum_i R_i/q_c(X_i,\theta_i)}{\sum_i 1/q_c(X_i,\theta_i)}
\label{eq:ipw-ratio}
\end{equation}
is the natural variant. It differs from \eqref{eq:ipw} by $O_p(n^{-1/2})$, reduces to the sample mean when $q_c\equiv1$, and is the form whose finite-sample bounds are machine-checked. With the design weight inserted, numerator and denominator both carry $w(X_i)/q_c(X_i,\theta_i)$, and the estimator is the ratio of the two weighted sums.

\subsection{The naive bias mechanism}
\label{sec:naive-bias}

The return fraction $\bar R=n^{-1}\sum_iR_i$, reported without weights, does not have expectation $p_{\mathrm{surf}}$. Averaging over the canopy model as well as the design, under surface-uniform sampling and constant cover $p\equiv p_0$, equation~\eqref{eq:detection} gives $\mathbb{E}[R_i\mid X_i]=p_0q_c(X_i,\theta_i)$, hence
\begin{equation}
\mathbb{E}[\bar R]=p_0\,\mathbb{E}_\mu[q_c].
\label{eq:naive-mean}
\end{equation}
Since $q_c\le1$, the naive fraction estimates $p_0\mathbb{E}_\mu[q_c]$, not $p_0$, and its bias $p_0(\mathbb{E}_\mu[q_c]-1)$ is negative. Without constant cover the same argument gives $\mathbb{E}[\bar R]=\mathbb{E}_\mu[p\,q_c]$, which equals $p_{\mathrm{surf}}$ only when $q_c$ is independent of $p$ or $p$ is constant. The bias is thus a property of the detection regime.

The simulations give the size of the effect and show that it is regime-specific. Under the phenomenological detection model of Section~\ref{sec:validation}, with $p_0=0.5$ and constant cover, the naive bias at $n=5000$ was $-0.413$ on the flat surface and $-0.400$ on the hill (Table~\ref{tab:detection-bias}). Read through \eqref{eq:naive-mean}, these imply $\mathbb{E}_\mu[q_c]\approx0.173$ and $\approx0.201$: on average a pulse had roughly one chance in six to one in five of returning a canopy label where canopy was present. In the caricature $q=0.15+0.85\rho^{-2}$, in which only slope matters, the naive fraction was biased by $-0.040$ at $n=500$, an order of magnitude less. Neither figure is a claim about a real scanner: they are two points on the map from detection regime to bias drawn by \eqref{eq:naive-mean}.

\subsection{Known detection and its limits}
\label{sec:identification}

Estimator \eqref{eq:ipw} presumes that $q_c$ is known, or at least correctly modelled, and that presumption is the main restriction here. Four assumptions are in force: (i) pulse locations are independent draws from the known design $\xi$; (ii) detection satisfies \eqref{eq:detection}; (iii) $q_c$ is known and strictly positive; (iv) $\rho$ is bounded and integrable on $D$, so that $\mu$ is a probability measure and $w$ is finite. Assumptions (i) and (iv) hold by construction in the simulations, and (ii) holds because the terrain-derived density proxy is independent of the realised canopy. Assumption (iii) is the one that does not transfer.

In a real acquisition $q_c$ is not identified from a single scan. It depends on incidence angle, range, occlusion, pulse loss and clumping, and clumping depends on the canopy structure the survey is trying to measure \citep{xu2026quantifying}. Scan angle alone changes cover estimates materially beyond about $12^\circ$ \citep{ma2017comparison}; pulse loss requires a loss-ratio correction \citep{liu2019improving}; slope distorts waveforms and return densities \citep{park2014application,li2023topography,hou2026slope}; occlusion makes the detection function platform-specific \citep{schneider2019quantifying,donager2021adjudicating}; and cover models transfer poorly across ecozones \citep{hopkinson2009testing}. A usable $q_c$ must be assembled from scan geometry, a canopy height model and a clumping model, and its errors propagate into $\hat p$ and into the test of Section~\ref{sec:global-test}. Misspecification is not a technicality: in an earlier pilot with the simpler slope-detection model, replacing $q_c$ by its mean raised the null rejection rate of a local scan from $0.07$ to $0.10$. We therefore present \eqref{eq:ipw} as a detection-corrected estimator under a stated detection model, not as an estimator robust to an unknown one. Weights estimated from a misspecified model can fail badly \citep{kang2007demystifying}; the detection weight is exposed to the same failure, and only the independence of the terrain-derived density proxy protects the simulation.

\subsection{What is checked}
\label{sec:ipw-checks}

The finite-sample algebra of \eqref{eq:ipw} and \eqref{eq:ipw-ratio} is machine-checked: nonnegativity; the bound $\hat p\le1/q_{\min}$ when $R_i\le1$ and $q_c\ge q_{\min}>0$; the unit-interval bound for the ratio form with binary returns; the no-detection reduction to the sample mean; the all-canopy and all-ground values; and additivity in the return indicators. What is not checked is any statement about the distribution of $\hat p$: equations~\eqref{eq:unbiased} and \eqref{eq:ipw-var} are ordinary expectation computations, and consistency is not claimed. Section~\ref{sec:lean} describes the formal core and its audit.

\section{The global lack-of-fit test}
\label{sec:global-test}

Section~\ref{sec:sampling} estimates how much of the surface is covered; this section asks whether its mean is constant. The two questions share a weighted empirical process, and the difficulty of the second lies in its null law, not in the statistic.

\subsection{Three nulls}
\label{sec:nulls}

Three nested hypotheses matter. Random labelling, $H_0^{\mathrm{random}}$, says that given the pulse locations the return indicators are exchangeable. Constant mean surface-area cover, $H_0^{\mathrm{const}}$, says that $\mathbb{E}[C(s)\mid s]=p_0$ for every $s$, spatial correlation in $C$ being allowed. Terrain independence, $H_0^{\mathrm{terrain}}$, says that $\mathbb{E}[C(s)\mid\mathrm{slope}(s)]=p_0$. The three are nested: random labelling implies constant mean cover, and constant mean cover implies terrain independence. The scientific hypothesis of interest is the middle one; the reported calibration targets the independent-return sub-null $H_{0,\mathrm{ind}}^{\mathrm{const}}$ defined next.

The target null and the calibrated null must be kept apart. Let $H_{0,\mathrm{ind}}^{\mathrm{const}}$ be $H_0^{\mathrm{const}}$ together with independent canopy returns, that is, $C(s)$ independent of $C(s')$ for $s\ne s'$. The parametric bootstrap of Section~\ref{sec:bootstrap} calibrates $H_{0,\mathrm{ind}}^{\mathrm{const}}$, not the broader $H_0^{\mathrm{const}}$: it draws independent Bernoulli labels and so reproduces the covariance of an independent-label process, not that of a correlated field. No claim of size $0.05$ is made for an arbitrary correlated constant-mean field; the correlated case is outside the calibration and is discussed in Section~\ref{sec:bootstrap}. Testing the mark law rather than the location law is itself established \citep{mrkvicka2012goodness}; the null there is a parametric mark model, whereas the null here is constancy of the mean against a terrain surface measure.

Two consequences follow. A rejection of $H_0^{\mathrm{const}}$ is not evidence of terrain dependence in particular: a compact hotspot departs from constancy with no slope gradient, and is reported as its own alternative. And a test calibrated for random labelling is not a test of constant mean cover; nor is a test calibrated for $H_{0,\mathrm{ind}}^{\mathrm{const}}$ automatically a test of the correlated $H_0^{\mathrm{const}}$.

\subsection{The statistic}
\label{sec:statistic}

Let $c\in\{0,1\}^2$ index a corner of the square, with $s\le_cx$ meaning $s_j\le x_j$ where $c_j=0$ and $s_j\ge x_j$ where $c_j=1$, coordinatewise. For $x\in D$ let $F_{\mu,c}(x)=\mu\{s:s\le_cx\}$ be the surface corner CDF. The weighted empirical canopy-return CDF at corner $c$ is
\begin{equation}
\hat F_c(x)=\frac{1}{n}\sum_{i=1}^{n}w_i\,\frac{R_i}{q_c(X_i,\theta_i)}\,\mathbf{1}\{X_i\le_cx\},
\qquad w_i=\frac{\rho(X_i)/Z}{g(X_i)},
\label{eq:ipw-edf}
\end{equation}
where $w_i\equiv1$ under surface-uniform sampling and $w_i=\rho(X_i)/Z$ under plane-uniform sampling, the case exercised by the plane arm of Section~\ref{sec:plane-results}. The bootstrap of Section~\ref{sec:bootstrap} recomputes $T^{*}$ with the same $w_i$, so the two arms differ only in the weights, not in the definition of $T$. The per-corner statistic is
\begin{equation}
T_c=\int_D\bigl(\hat F_c(x)-\hat p_w\,F_{\mu,c}(x)\bigr)^2\,\mu(dx),
\label{eq:per-corner}
\end{equation}
with $\hat p_w=n^{-1}\sum_iw_iR_i/q_c(X_i,\theta_i)$ the weighted analogue of \eqref{eq:ipw}, which reduces to \eqref{eq:ipw} when $w\equiv1$, and the global statistic is the average over corners,
\begin{equation}
T=\frac14\sum_{c}T_c .
\label{eq:global}
\end{equation}
The integral in \eqref{eq:per-corner} is with respect to the surface measure, so corner boxes are weighted by surface area and the statistic is the $L^2(\mu)$ distance appropriate to the estimand. The corner average follows \citet{zimmerman1993bivariate}, who introduced it so that no corner is privileged; a related Cram\'er--von Mises-type statistic addresses separability of marks and locations \citep{schoenberg2004testing}. Weighted Cram\'er--von Mises criteria are studied by \citet{ozturk1997generalised}, and the effect of estimating the centre from the same sample---which separates \eqref{eq:per-corner} from a fixed-parameter statistic---was studied by \citet{martynov2011weighted}. Neither covers a graph surface or a detection-weighted process.

\subsection{The mean is right, the null law is not}
\label{sec:mean-right}

Under $H_0^{\mathrm{const}}$ the weighted process is correctly centred. For every $x$,
\begin{equation}
\begin{aligned}
\mathbb{E}[\hat F_c(x)]
&=\frac{1}{n}\sum_{i=1}^{n}\mathbb{E}_\xi\!\left[w_i\,\frac{\mathbf{1}\{X_i\le_cx\}}{q_c(X_i,\theta_i)}\,\mathbb{E}\{R_i\mid X_i\}\right]\\
&=\frac{p_0}{n}\sum_{i=1}^{n}\mathbb{E}_\xi\bigl[w_i\,\mathbf{1}\{X_i\le_cx\}\bigr]=p_0F_{\mu,c}(x),
\end{aligned}
\label{eq:centred}
\end{equation}
using $\mathbb{E}\{R_i\mid X_i\}=p_0q_c(X_i,\theta_i)$ under $H_0^{\mathrm{const}}$ for the second equality, and $\mathbb{E}_\xi[w_i\,\mathbf{1}\{X_i\le_cx\}]=\mu\{s:s\le_cx\}$, which follows from $wg=\rho/Z$, for the third. This holds whatever the spatial correlation of $C$, because \eqref{eq:ipw-edf} is a sum of terms, each depending on one pulse, and only the marginal law of $R_i$ enters. Correlation changes the covariance of $\hat F_c$ across $x$, and therefore the distribution of $T$, but not its mean. The statistic needs no new centre; it needs the law of a weighted empirical process under $H_0^{\mathrm{const}}$ with spatial correlation, which is not available in closed form. Asymptotic goodness-of-fit theory for the Palm mark distribution with correlated marks exists \citep{heinrich2014asymptotic}; it does not cover the weighted surface process, whose reference measure and weights both depend on the terrain. Calibration is the whole problem.

\subsection{Parametric bootstrap}
\label{sec:bootstrap}

The reported calibration is a parametric bootstrap under $H_{0,\mathrm{ind}}^{\mathrm{const}}$: constant mean cover with independent canopy returns. Holding the observed locations, detection probabilities and design weights fixed, draw
\begin{equation}
R_i^{*}\sim\mathrm{Bernoulli}\bigl(p_0\,q_c(X_i,\theta_i)\bigr),\qquad i=1,\dots,n,
\label{eq:bootstrap}
\end{equation}
independently, recompute $T^{*}$ from \eqref{eq:ipw-edf}--\eqref{eq:global} with the same weights $w_i$, and repeat $B$ times; with $B=99$ the p-value is $\bigl(1+\#\{T_b^{*}\ge T\}\bigr)/(B+1)$. Because $p_0$ is fixed at $0.5$ in every reported run, this is a simple-null bootstrap; a composite-null version would plug in $\hat p_w$ and account for its estimation error, which we have not done.

The scheme is model-based twice over: it conditions on the detection probabilities and draws independent labels. It is a null for $H_{0,\mathrm{ind}}^{\mathrm{const}}$, not for the correlated $H_0^{\mathrm{const}}$. A strongly correlated constant-mean field would be rejected about as readily as a non-constant mean, so the test does not separate lack of constancy from unmodelled correlation. Detection and correlation were not combined in one study.

\subsection{Why label permutation is the wrong null}
\label{sec:permutation}

Label permutation reallocates the observed return indicators among the pulse locations and recomputes $T$. It is exact for $H_0^{\mathrm{random}}$, under which the observed assignment is one of the equally likely permutations. It is not a test of $H_0^{\mathrm{const}}$. Permutation makes the label law location-independent by construction, so the null has the covariance of a location-independent label process, whereas a constant-mean field may be strongly clustered. Positive correlation concentrates labels in patches, so $\hat F_c$ moves in large increments at patch boundaries and $T$ acquires a heavier upper tail than the permutation null admits. The rejection probability therefore depends on the correlation, and at $n=500$ with correlation length $0.1$ it reached $1.000$ on all three surfaces, against a nominal $0.05$.

\citet{grabarnik2011correct} show that a deviation test is a test of the hypothesis its scheme randomises; random-shift and block schemes restore size under autocorrelation \citep{mrkvicka2020revisiting,radersma2015new}, and local random-labelling tests make the same distinction regionally \citep{dangelo2023local}. In the ideal-sampling study a toroidal shift gave $0.040$--$0.055$ and a thresholded-field bootstrap $0.055$--$0.060$. The failure is thus the scheme's, not the statistic's: the same $T$ is calibrated once the null has the right covariance.

Under detection there is a second defect. In the surface-uniform case, permutation holds the observed return count fixed and redistributes it over locations whose weights $1/q_c$ differ, so the permuted mean of the weighted EDF is $\bar R\,n^{-1}\sum_i\mathbf{1}\{X_i\le_cx\}/q_c(X_i,\theta_i)$, not $p_0F_{\mu,c}(x)$: when $q_c$ varies, the null is mis-centred as well as mis-scaled.

\subsection{Reductions to existing statistics}
\label{sec:reductions}

Three exact reductions connect $T$ to simpler statistics, and each is machine-checked. With $q_c\equiv1$ and $w\equiv1$ the weighted EDF \eqref{eq:ipw-edf} becomes the label-weighted EDF and $T$ becomes the cover-label CvM (\texttt{ipwSurfaceCvM\_eq\_coverCvM\_of\_q\_one}). With $R\equiv1$, $q_c\equiv1$ and $w\equiv1$ it becomes the location-only surface CvM (\texttt{ipwSurfaceCvM\_one}). On the flat graph, where $\mu$ is Lebesgue measure on the square, that statistic equals the corner-averaged statistic $\bar\omega^2_D$ (\texttt{surfaceCvM\_flatSurface}), where $\bar\omega^2_D$ is the corner-averaged Cram\'er--von Mises statistic for spatial randomness on the unit square (the $d=2$ statistic of \citet{zimmerman1993bivariate}); \texttt{omegaBarSqD} is its Lean name. The chain is a continuity check: on a flat surface without detection the new statistic is the same functional as the old one. It says nothing about critical values, and no null law is read off it; no general non-flat reduction is claimed.

\subsection{Computed statistic and operating range}
\label{sec:operating-range}

The integral in \eqref{eq:per-corner} is computed as an average over $m_{\mathrm{ref}}=150$ reference points drawn from $\mu$ and redrawn in each replication, so the computed statistic is a Monte Carlo approximation of $T$. Observed and bootstrap statistics share the reference sample within a replication, so the error largely cancels, though it adds variability the calibration does not model.

Figure~\ref{fig:pulse-density} summarises the operating range; Table~\ref{tab:global-ipw-cvm} reports the operating characteristics at $\alpha=0.05$, with $B=99$ and 100 replications per cell. On the hill, size was $0.03$, $0.05$ and $0.04$ at $n=500$, $2000$ and $5000$, and the flat cells lay between $0.02$ and $0.07$; all are compatible with a nominal $0.05$ at 100 replications in these independent-return cells. Power against the broad slope-dependent alternative on the hill was $0.03$ at $n=500$, $0.26$ at $n=2000$ and $0.79$ at $n=5000$ ($0.70$--$0.87$): no power beyond the nominal level at 500 pulses, and 79 rejections in 100 replications at 5000. Under plane-uniform sampling, the $w_i=\rho(X_i)/Z$ case, the $n=5000$ cell gave size $0.04$ and power $0.85$ ($0.77$--$0.91$). The plane arm shows that the geometric weight can be carried without breaking calibration. Against the compact hotspot, power was $0.12$ ($0.06$--$0.20$) on the hill and $0.13$ under the plane design.

Two cautions follow. There is no design-based asymptotic null law for $T$: the calibration is the parametric bootstrap of Section~\ref{sec:bootstrap}, and its validity is conditional on the detection model, on independent returns (that is, on $H_{0,\mathrm{ind}}^{\mathrm{const}}$) and on the simple null $p_0=0.5$. And $0.79$ is a single synthetic cell---one surface, one alternative strength ($\beta=0.25$), one pulse density, 100 replications---not an asymptotic guarantee. It establishes that the test is not structurally powerless against broad terrain dependence at high pulse density, not a rate, a threshold, or a hotspot detector. Against the compact hotspot the test is nearly powerless, and no local detector is claimed here.

\section{The Lean 4 formal core}\label{sec:lean}

The objects defined in the preceding sections are formalised in a Lean~4 development
\citep{demoura2021lean4} built on Mathlib \citep{mathlib2020lean}. The purpose is narrow: a
formalisation cannot certify a test, and nothing below is evidence that the weighted statistic
has the size or power reported in Section~\ref{sec:validation}. It fixes the definitions and
checks the algebraic and measure-theoretic steps that relate them, so that later asymptotic work
cannot silently substitute a different functional. Formalised probability already contains limit
theorems---\citet{avigad2017formally} verify the central limit theorem in Isabelle---which
marks where this core stops: it checks the objects and reductions, not the null law.

\paragraph{Geometry} A terrain surface is a \texttt{GraphSurface}: a continuously
differentiable height function $f$ over the parameter square. Its area density is
\[
  \rho(x)=\Bigl(1+\sum_{i=1}^{2}\bigl(\partial_i f(x)\bigr)^{2}\Bigr)^{1/2},
\]
which is \texttt{areaDensity}. The sum-of-squares form is deliberate: the Pi-type norm on
$\mathrm{Fin}\ 2\to\mathbb{R}$ is the sup norm, so the operator norm of the derivative is not the
Euclidean gradient norm. The module checks that \texttt{areaDensity} is strictly positive,
continuous and bounded below by one
(\texttt{areaDensity\_pos}, \texttt{areaDensity\_continuous}, \texttt{one\_le\_areaDensity});
that the area-weighted measure \texttt{surfaceMeasure} has finite and nonzero total mass
(\texttt{surfaceArea\_ne\_top}, \texttt{surfaceArea\_ne\_zero}); and that
\texttt{normalizedSurfaceMeasure} is a probability measure
(\texttt{isProbabilityMeasure\_normalizedSurfaceMeasure}). On the flat graph the density is one
and the surface measure is Lebesgue measure on the square (\texttt{surfaceMeasure\_flat},
\texttt{normalizedSurfaceMeasure\_flat}).

\paragraph{Statistics and reductions} The surface statistic integrates the squared difference
between the empirical corner CDF and the surface corner CDF against $\mu$, and averages over
the four corners; \texttt{surfaceCvM\_nonneg} records nonnegativity. The cover-label and
weighted versions replace it by a label average and by an inverse-probability-weighted return
average. The formal core fixes the surface-uniform case $w\equiv1$; the general design weight
$w_i$ of \eqref{eq:ipw-edf} is paper-level and is not formalized. Write $X=(X_1,\dots,X_n)$ for the sample and $\mu_{\mathrm{flat}}$ for the surface measure of the flat graph, which is Lebesgue measure on $D$. Each reduction is a separate theorem:
\[
  \mathtt{surfaceCvM}(\mu_{\mathrm{flat}},X)
    =\mathtt{omegaBarSqD}(X),
\]
\[
  \mathtt{coverCvM}(\mu,\mathbf{1},X)
    =\mathtt{surfaceCvM}(\mu,X),
\]
\[
  \mathtt{ipwSurfaceCvM}(\mu,R,\mathbf{1},X)
    =\mathtt{coverCvM}(\mu,R,X),
\]
\[
  \mathtt{ipwSurfaceCvM}(\mu,\mathbf{1},\mathbf{1},X)
    =\mathtt{surfaceCvM}(\mu,X).
\]
The first is \texttt{surfaceCvM\_flatSurface}, the second \texttt{coverCvM\_one}, the third
\path{ipwSurfaceCvM_eq_coverCvM_of_q_one}, and the fourth \texttt{ipwSurfaceCvM\_one},
which composes the second and third; the second and fourth need a nonempty sample. Read as a
chain, they say
that the weighted statistic with no detection and unit labels is the location-only surface
statistic, and that on the flat graph this is the existing corner-averaged statistic.
They are a continuity check, not a null law: they say nothing about critical values.

\paragraph{Permutation and local scan} Two combinatorial facts are recorded: simultaneous
relabeling of labels and locations leaves the label-weighted statistic unchanged
(\texttt{coverCvM\_comp\_perm}), and permuting the observed labels leaves its permutation orbit
unchanged (\texttt{coverCvMPermOrbit\_relabel}). The second is the combinatorial invariance
behind a permutation calibration; it is not a statement that the permutation $p$-value is
valid, as the module docstring states. The local scan is defined over a finite
family of measurable patches, with a conditional corner CDF and a local CvM. The module checks
whole-patch recovery (\texttt{localCoverCvM\_whole}), scan dominance when the whole square
belongs to the family (\texttt{coverCvM\_le\_surfaceScan\_of\_whole}), and a noncentrality
ingredient (\texttt{localCoverShift\_noncentrality\_pos}). It supplies no null law for the
maximum and no multiple-testing adjustment, which is why Section~\ref{sec:local} reports a
negative result rather than a test.

\paragraph{Modules} Table~\ref{tab:lean-core} tabulates the declarations and audit roles.
\begin{itemize}
  \item \texttt{SurfaceMeasure}: graph, area density, area measure; positivity,
        continuity, $\rho\ge1$, finiteness, normalisation, and the flat-graph identification.
  \item \texttt{SurfaceCvM}: surface corner CDF and corner-averaged squared $L^{2}(\mu)$
        distance; nonnegativity and the flat reduction to \texttt{omegaBarSqD}.
  \item \texttt{SurfacePermutation}: cover labels, fitted cover, label-weighted CvM and
        permutation orbit; nonnegativity, simultaneous-relabeling invariance, orbit invariance
        and the all-canopy reduction.
  \item \texttt{SurfaceScan}: patches, conditional corner CDF, local CvM, maximum over a
        finite family; whole-patch recovery, scan dominance, noncentrality.
  \item \texttt{SurfaceCoverIPW}: inverse-probability-weighted mean and ratio forms;
        nonnegativity, the $1/q_{\min}$ bound, the unit-interval bound, additivity in the
        returns, and the no-detection, all-canopy and all-ground identities.
  \item \texttt{SurfaceIPWCvM}: weighted empirical CDF and corner-averaged weighted CvM;
        nonnegativity, the no-detection reduction to \texttt{coverCvM}, and the
        all-canopy/no-detection reduction to \texttt{surfaceCvM}.
\end{itemize}

\paragraph{Audit and scope} The headline theorems, together with the three area-density
regularity lemmas, are registered in \texttt{verify/AxiomAudit.lean}, and the audit passes:
each prints only the standard three axioms, \texttt{[propext, Classical.choice, Quot.sound]};
the audit command is in Appendix~\ref{app:replication}. We do not quote a count of audited
declarations, because it drifts as the library grows. What is checked is the Euclidean graph
density and its regularity; finiteness, positivity and normalisation of the surface measure;
nonnegativity of the three statistics; the finite-sample bounds on the weighted estimator
(\texttt{ipwCover\_nonneg}, \texttt{ipwCover\_le\_inv\_qmin},
\texttt{ipwCoverRatio\_mem\_unit}); the no-detection, all-canopy and all-ground reductions;
whole-patch recovery and scan dominance; and the flat-surface identification with
\texttt{omegaBarSqD}. What is not checked is any null distribution, critical value or asymptotic
theory; design-unbiasedness of the weighted estimator; consistency; change-of-variables
invariance of the surface measure; mesh or triangle surfaces; and the multiple-testing
adjustment for the scan. No general non-flat reduction to a product-corner statistic is claimed. The correct
summary is that the definitions and their reductions are machine-checked; the method is not
verified.

\begin{table}[t]
\centering
\caption{Declarations in \texttt{lean/SteinGate/CvM/Surface*.lean} that the paper relies
on. Every theorem listed is registered in \texttt{verify/AxiomAudit.lean} and prints only
\texttt{[propext, Classical.choice, Quot.sound]}; definitions are included for orientation
and are not axiom-audited. The table reports what is checked: definitions, regularity,
bounds and exact reductions. No null law, unbiasedness, consistency or
p-value validity statement is among them.}
\label{tab:lean-core}
\footnotesize
\begin{tabular}{@{}>{\raggedright\arraybackslash}p{0.24\textwidth}>{\raggedright\arraybackslash}p{0.44\textwidth}>{\raggedright\arraybackslash}p{0.24\textwidth}@{}}
\toprule
Module & Headline declarations & What is checked \\
\midrule
\texttt{SurfaceMeasure.lean}
  & \texttt{areaDensity}, \texttt{surfaceMeasure}, \texttt{surfaceArea},
    \texttt{normalizedSurfaceMeasure}, \texttt{surfaceMeasure\_flat},
    \texttt{normalizedSurfaceMeasure\_flat},
    \texttt{areaDensity\_pos}, \texttt{areaDensity\_continuous},
    \texttt{one\_le\_areaDensity}
  & Euclidean graph area element $\rho=\sqrt{1+\lVert\nabla f\rVert^{2}}$; the
    area-weighted measure; finite, positive, normalised; exact flat reduction \\
\addlinespace
\texttt{SurfaceCvM.lean}
  & \texttt{surfaceCornerCdf}, \texttt{surfaceCvM},
    \texttt{surfaceCvM\_nonneg}, \texttt{surfaceCvM\_flat},
    \texttt{surfaceCvM\_flatSurface}
  & Corner-averaged squared $L^{2}(\mu)$ distance; nonnegativity; exact flat-surface
    reduction to \texttt{omegaBarSqD} \\
\addlinespace
\texttt{SurfacePermutation.lean}
  & \texttt{coverMean}, \texttt{coverEdf}, \texttt{coverCvM},
    \texttt{coverCvM\_comp\_perm}, \texttt{coverCvMPermOrbit\_relabel},
    \texttt{coverCvM\_one}
  & Label-weighted CvM; simultaneous-relabelling invariance; the permutation-orbit
    identity; all-canopy reduction \\
\addlinespace
\texttt{SurfaceScan.lean}
  & \texttt{SurfacePatch}, \texttt{localSurfaceCornerCdf},
    \texttt{localCoverCvM}, \texttt{surfaceScan},
    \texttt{coverCvM\_le\_surfaceScan\_of\_whole},
    \texttt{localCoverShift\_noncentrality\_pos}
  & Conditional patch CDF, local CvM, maximum over a finite patch family; the scan
    dominates the global statistic when the whole square is a patch; a noncentrality
    ingredient \\
\addlinespace
\texttt{SurfaceCoverIPW.lean}
  & \texttt{ipwCover}, \texttt{ipwCoverRatio}, \texttt{ipwCover\_nonneg},
    \texttt{ipwCover\_le\_inv\_qmin}, \texttt{ipwCoverRatio\_mem\_unit},
    \texttt{ipwCover\_eq\_mean\_of\_q\_one}
  & Horvitz--Thompson mean and ratio form; nonnegativity; the $1/q_{\min}$ bound; the
    ratio lies in $[0,1]$; no-detection reduction to the sample mean \\
\addlinespace
\texttt{SurfaceIPWCvM.lean}
  & \texttt{ipwEdf}, \texttt{ipwSurfaceCvMAtCorner}, \texttt{ipwSurfaceCvM},
    \texttt{ipwSurfaceCvM\_nonneg},
    \texttt{ipwSurfaceCvM\_eq\_coverCvM\_of\_q\_one}, \texttt{ipwSurfaceCvM\_one}
  & Surface-uniform ($w\equiv1$) weighted empirical canopy-return CDF; corner-averaged weighted CvM; nonnegativity;
    exact no-detection and all-canopy reductions \\
\bottomrule
\end{tabular}
\end{table}

\section{Synthetic validation}\label{sec:validation}

\subsection{Design}\label{sec:design}

All experiments are on the parameter square $D=[0,1]^{2}$, with three surfaces: a flat plane,
$f=0$; a constant slope, $f(u,v)=0.5u$, whose area ratio is $1.1180$; and a Gaussian hill,
\[
  f(u,v)=0.45\exp\!\left(-\frac{(u-0.5)^{2}+(v-0.5)^{2}}{2(0.18)^{2}}\right),
\]
whose area ratio is $1.2477$. Each surface is crossed with three cover maps: constant
$p_{0}=0.5$; a central-disc hotspot of $+0.2$ over about $10\%$ of the planimetric area; and a
slope-dependent map $p(u,v)=0.5+0.25(\mathrm{slope}(u,v)-\overline{\mathrm{slope}})$, clipped to
$[0,1]$; we write $\beta=0.25$ for the coefficient. Canopy labels are independent Bernoulli draws with probability $p(X_{i})$; the
correlated-label study belongs to the calibration discussion, not here.

Pulses are emitted at locations $X_{i}$, and the observed outcome is the canopy return indicator
$R_{i}=C(X_{i})\mathbf{1}\{\text{canopy return recorded}\}$, so that
$R_{i}\mid X_{i}\sim\mathrm{Bernoulli}(p(X_{i})q_{c}(X_{i}))$. The detection probability is the
phenomenological product
\[
  q_{c}
  =\underbrace{q_{0}\,\cos\theta\,e^{-r/R_{0}}}_{q_{\mathrm{surf}}}
   \;\underbrace{\left(1-e^{-k_{\mathrm{lai}}d/\cos\theta}\right)}_{p_{\mathrm{int}}}\;\eta_{c},
  \qquad
  q_{g}=\eta_{g}\,q_{\mathrm{surf}}\,(1-p_{\mathrm{int}}),
\]
with $q_{0}=0.90$, $R_{0}=2.0$, $k_{\mathrm{lai}}=1.5$, $\eta_{c}=0.90$ and
$\eta_{g}=0.80$. The flight line runs along $v$ at $u=0.5$ and height $H=1$, in units of the
unit square. For a pulse at $(u,v)$, the cross-track offset is $du=u-0.5$ and the slant range is
$r=(H^{2}+du^{2})^{1/2}$, measured to the planimetric position. The laser direction is
$(du,0,-H)/r$ and the surface normal is $(-f_{u},-f_{v},1)/\rho$; the incidence cosine
$\cos\theta$ is the absolute dot product of the two, clipped to $[0.05,1]$. The field $d$ is not
an independent canopy density: it is the normalised terrain-derived proxy
\[
  d=0.3+0.6\,\frac{\rho-\rho_{\min}}{\rho_{\max}-\rho_{\min}}\in[0.3,0.9],
\]
a deterministic function of terrain slope, independent of the realised binary canopy. All
constants are in these normalised units; no physical (metre) scale is implied. Since
$q_{c}+q_{g}=q_{\mathrm{surf}}(0.8+0.1p_{\mathrm{int}})\le0.81<1$, the renormalisation guard in
the code is inactive in these runs, and pulses with neither return are simply lost. The
components follow the applied literature on scan
angle, pulse loss and clumping \citep{ma2017comparison,liu2019improving,xu2026quantifying};
parameter values are in the results file.

Two designs are used. In the surface-uniform design $X_{i}$ is drawn from the normalised
surface measure, so the area density cancels and the weight is $1/q_{c}(X_{i})$ alone. In the
plane-uniform design $X_{i}$ is uniform on $D$ and the weight is $\rho(X_{i})/Z$ divided by
$q_{c}(X_{i})$, with $Z=\int_{D}\rho$; this is closer to an airborne design than
surface-uniform sampling, and it is the only design that exercises the geometric weight. The global runs use the hill at
$n\in\{500,2000,5000\}$ pulses, with the flat surface as a null control, 100 replications,
$B=99$ bootstrap draws, a reference sample of $m_{\mathrm{ref}}=150$ points and a
$350\times350$ grid. Table~\ref{tab:designs} records the settings.

\begin{table}[t]
\centering
\caption{Recorded settings of the five synthetic studies. R0 has both an ideal-sampling
(no-detection) arm and a detection arm, the latter using the caricature model
$q=0.15+0.85\rho^{-2}$; R1--R4 use the phenomenological canopy-return model of
\texttt{simulate\_realistic\_detection.py}. The studies are stored in
\texttt{results.json} (R0), \texttt{realistic\_detection\_results.json} (R1),
\texttt{global\_realistic\_results.json} (R2), \texttt{multiscale\_rescue\_results.json}
(R3) and \texttt{global\_pulse\_density.json} (R4). Entries marked $\dagger$ are read
from the generating script because the JSON configuration block does not record them.
R0's $B=199$ is a permutation count; R1--R4 use $B=99$ bootstrap draws.}
\label{tab:designs}
\footnotesize
\setlength{\tabcolsep}{3pt}
\begin{tabular}{@{}llrrrrrr@{}}
\toprule
Study & Surfaces & $n$ & Reps & $B$ & $m_{\mathrm{ref}}$ & Grid & Seed \\
\midrule
R0 & flat, slope, hill & 500        & 200 & 199 & ---      & 500 & 20261102 \\
R1 & flat, hill  & 500        & 100 & 99  & 60       & 350 & 20261103 \\
R2 & flat, hill  & 5000       & 100 & 99  & 150      & 350 & 20261103 \\
R3 & hill        & 1000, 2000 & 40  & 50  & 40       & 350$^\dagger$ & 20261103$^\dagger$ \\
R4 & flat, hill  & 500, 2000, 5000 & 100 & 99 & 150 & 350 & 20261103 \\
\bottomrule
\end{tabular}
\end{table}

\subsection{The estimand on the three surfaces}\label{sec:estimand-results}

Table~\ref{tab:estimand} reports the two cover fractions. On the flat surface both are $0.5000$
and the area ratio is $1.0000$. On the constant slope the ratio is $1.1180$, yet under the
slope-dependent map both fractions are $0.5000$ and the gap is zero: a non-unit area ratio alone
does not change the estimand. On the hill the ratio is $1.2477$; constant cover gives
$0.5000$ against $0.5000$, and the hotspot gives $0.5258$ against $0.5200$ (a gap of $+0.0058$).
The sign is not an error: the disc radius $\sqrt{0.1/\pi}\approx0.178$ coincides with the steep
annulus of the Gaussian hill (scale $0.18$), so the disc's mean area density is about $1.61$
against the global $1.2477$, and the covariance is positive. Slope-dependent cover gives
$0.5276$ against $0.5000$ (a gap of $+0.0276$). The covariance
identity predicts a gap of
$\mathrm{Cov}_{\mathrm{planar}}(C,\rho)/\mathbb{E}_{\mathrm{planar}}[\rho]
=0.03448/1.24770=0.02763$, matching the recorded difference; the identity is exact, and the constant-slope control is the independent check.
The magnitude is modest---$2.8$ percentage points---and the hill is a gentle test case (height
$0.45$, width $0.18$). The claim is the identity, not the size of the gap.

\begin{table}[t]
\centering
\caption{Terrain-surface-area cover \(p_{\mathrm{surf}}\) against planimetric cover
\(p_{\mathrm{planar}}\), with the mean area density
\(\bar\rho=A_{\mathrm{surf}}/A_{\mathrm{plan}}\). Values are exact grid evaluations of the
analytic surfaces and the analytic cover maps on a \(500\times500\) grid; they are not
Monte Carlo estimates and carry no simulation error. The cover maps are constant
(\(p_{0}=0.5\)), central hotspot (\(+0.2\) on a disc of about 10\,\% of the planimetric
area) and slope-dependent (\(p(u,v)=0.5+0.25\,(\mathrm{slope}(u,v)-\overline{\mathrm{slope}})\), clipped to \([0,1]\)).}
\label{tab:estimand}
\begin{tabular}{@{}llrrrr@{}}
\toprule
Surface & Cover map & $p_{\mathrm{surf}}$ & $p_{\mathrm{planar}}$ & Difference & $\bar\rho$ \\
\midrule
Flat            & constant        & 0.5000 & 0.5000 & $0.0000$ & 1.0000 \\
Flat            & hotspot         & 0.5200 & 0.5200 & $0.0000$ & 1.0000 \\
Flat            & slope-dependent & 0.5000 & 0.5000 & $0.0000$ & 1.0000 \\
\addlinespace
Constant slope  & constant        & 0.5000 & 0.5000 & $0.0000$ & 1.1180 \\
Constant slope  & hotspot         & 0.5200 & 0.5200 & $0.0000$ & 1.1180 \\
Constant slope  & slope-dependent & 0.5000 & 0.5000 & $0.0000$ & 1.1180 \\
\addlinespace
Gaussian hill   & constant        & 0.5000 & 0.5000 & $0.0000$ & 1.2477 \\
Gaussian hill   & hotspot         & 0.5258 & 0.5200 & $+0.0058$ & 1.2477 \\
Gaussian hill   & slope-dependent & 0.5276 & 0.5000 & $+0.0276$ & 1.2477 \\
\bottomrule
\end{tabular}
\end{table}

\subsection{Detection bias}\label{sec:bias-results}

Table~\ref{tab:detection-bias} separates the two detection regimes. In the simple regime
$q(s)=0.15+0.85\rho(s)^{-2}$, a nadir-laser caricature in which steep terrain is undersampled,
the naive mean on the hill under slope-dependent cover at $n=500$ has bias $-0.0401$ and RMSE
$0.0449$, against $-0.0037$ and $0.0225$ for the ideal sample mean and $+0.0002$ and $0.0209$
for the ratio estimator with known $q$. In the realistic regime the naive canopy-return
fraction is badly biased: at $n=500$ it is $-0.4149$ on the flat surface and $-0.3994$ on the
hill, and at $n=5000$ it is $-0.4134$ and $-0.3997$. The inverse-probability-weighted estimator, with the
canopy detection probability as the inclusion probability \citep{horvitz1952generalization},
has bias $-0.0078$ and $+0.0010$ at $n=500$, and $-0.0002$ and $-0.0009$ at $n=5000$; across the
six cells of the $n=5000$ surface design its largest absolute bias is $0.003$.

The mechanism is visible without simulation. Under constant cover,
\[
  \mathbb{E}[\bar R]
  =\mathbb{E}_{\mu}[p(X)q_{c}(X)]
  =p_{0}\,\mathbb{E}_{\mu}[q_{c}],
\]
so the naive fraction estimates $p_{0}$ times the mean detection probability, and the bias
$p_{0}(\mathbb{E}_{\mu}[q_{c}]-1)$ is a property of the detection regime rather than an
accident of the sample. Inverting the recorded constant-cover biases gives an implied
$\mathbb{E}_{\mu}[q_{c}]\approx0.17$ on the flat surface and $\approx0.20$ on the hill; this is
derived, not recorded. The magnitude $-0.40$ is therefore specific to the
low-return regime simulated here and is not a claim about operational scanners.

\begin{table}[t]
\centering
\caption{Detection bias of the naive canopy-return fraction and of the
inverse-probability-weighted estimator. \textbf{(A)} Caricature model
\(q(s)=0.15+0.85\rho(s)^{-2}\) on the hill with slope-dependent cover, target
\(p_{\mathrm{surf}}=0.5276\); 200 replications, \(B=199\). \textbf{(B)}
Phenomenological canopy-return model (incidence angle, range, canopy interception, pulse
loss) under constant cover \(p_{0}=0.5\); 100 replications, \(B=99\), surface and plane
designs. Both panels report the mean of \((\hat p-p_{\mathrm{surf}})\) over replications;
RMSE is the root mean squared error over the same replications. All values are Monte Carlo
estimates from a synthetic pilot.}
\label{tab:detection-bias}
\begin{tabular}{@{}lrr@{}}
\toprule
\multicolumn{3}{@{}l}{\textbf{(A) Caricature slope-detection model, hill, slope-dependent cover}} \\
\addlinespace
Estimator & Bias & RMSE \\
\midrule
Ideal sample mean (no detection)      & $-0.0037$ & 0.0225 \\
Naive mean under detection            & $-0.0401$ & 0.0449 \\
Ratio estimator, known $q$            & $+0.0002$ & 0.0209 \\
\bottomrule
\end{tabular}\par\vspace{8pt}
\begin{tabular}{@{}llrrr@{}}
\toprule
\multicolumn{5}{@{}l}{\textbf{(B) Phenomenological canopy-return model, constant cover}} \\
\addlinespace
Surface & Design & $n$ & Naive bias & IPW bias \\
\midrule
Hill & surface & 500  & $-0.3994$ & $+0.0010$ \\
Hill & surface & 2000 & $-0.3999$ & $-0.0038$ \\
Hill & surface & 5000 & $-0.3997$ & $-0.0009$ \\
Hill & plane   & 5000 & $-0.4012$ & $+0.0023$ \\
\addlinespace
Flat & surface & 500  & $-0.4149$ & $-0.0078$ \\
Flat & surface & 2000 & $-0.4142$ & $-0.0053$ \\
Flat & surface & 5000 & $-0.4134$ & $-0.0002$ \\
Flat & plane   & 5000 & $-0.4132$ & $+0.0007$ \\
\bottomrule
\end{tabular}
\end{table}

\subsection{Size of the global test}\label{sec:size-results}

At $n=5000$ on the hill the parametric bootstrap holds the size at $0.04$ (95\% Clopper--Pearson
interval $0.011$--$0.099$); on the flat surface, where the
null holds by construction, it is $0.02$ ($0.002$--$0.070$). The flat slope-dependent row is
also a null, because $p(u,v)$ is constant on a flat surface, and it gives $0.03$
($0.006$--$0.085$); it is a free check that the test does not reject a constant field. At the
lower densities the hill gives $0.03$ at $n=500$ and $0.05$ at $n=2000$, and the flat surface
gives $0.07$ and $0.05$. These rates are consistent with a nominal $5\%$ level within binomial
error; we do not read the ordering as a trend.

\subsection{Power against broad terrain dependence}\label{sec:power-results}

\begin{figure}[t]
\centering
\includegraphics[width=\textwidth]{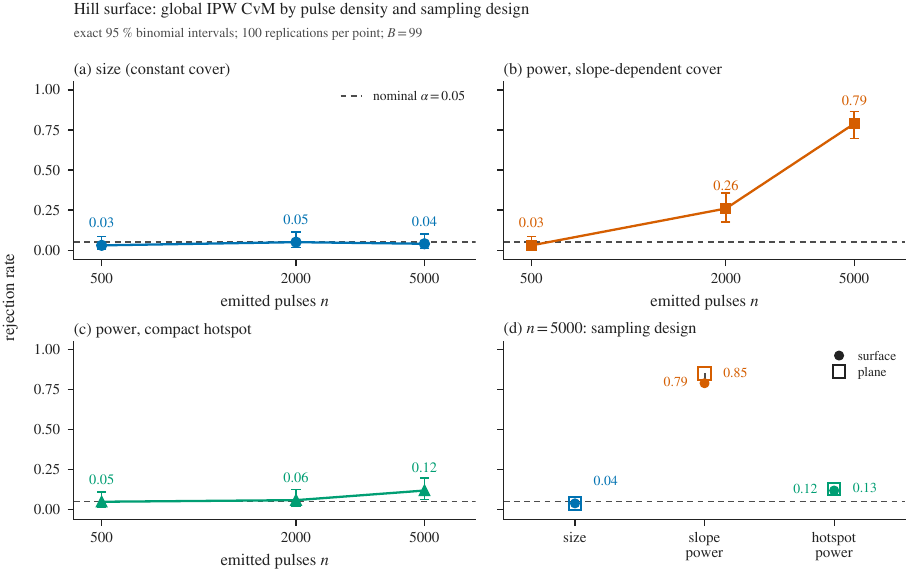}
\caption{Global corner-averaged IPW CvM on the hill, rejection rate at \(\alpha=0.05\)
against pulse density and sampling design. Panels (a)--(c): surface design at
\(n=500,2000,5000\) for the size of the test under constant cover, power against
slope-dependent cover, and power against a compact hotspot. Panel (d): the same three
quantities at \(n=5000\) under the surface design (filled circles) and the plane design
(open squares), where pulse locations are uniform on the parameter square and the weights
carry the extra factor \(\rho(X)/Z\). All points are Monte Carlo estimates from
100 replications with \(B=99\) parametric-bootstrap draws; bars are exact
Clopper--Pearson 95\,\% intervals. The dashed line is the nominal level. Synthetic pilot,
not field-validated.}
\label{fig:pulse-density}
\end{figure}

Figure~\ref{fig:pulse-density} summarises the density and design comparison, and
Table~\ref{tab:global-ipw-cvm} reports the three recorded densities. On the hill, power against the
slope-dependent alternative with $\beta=0.25$ is $0.03$ at $n=500$, $0.26$ at $n=2000$ and
$0.79$ at $n=5000$ ($0.697$--$0.865$); the last is the principal power result. Against the
compact hotspot the corresponding rates are $0.05$, $0.06$ and $0.12$
($0.064$--$0.200$), as expected of an integrated $L^{2}$ distance against a departure confined to
a tenth of the area. The three densities are reported as three
points, not as an estimated power curve: each rate has 100 replications, only one surface
geometry and one alternative strength are studied, and no asymptotic rate is claimed.

The naive column of Table~\ref{tab:global-ipw-cvm} is not a competitor. Its rejections refer to the
canopy-return fraction $\mathbb{E}[pq_{c}]$, not to $p_{\mathrm{surf}}$, and its bias on the hill
alternatives is about $-0.42$. It is calibrated here by the same parametric bootstrap, so its
near-nominal size is not evidence that label permutation is valid; under correlated cover the
permutation scheme has size $1.00$ in the ideal-sampling study. On the hill slope alternative
its rejection rate is $1.00$ at $n=5000$, but that is a rejection of the wrong null about the
wrong quantity, and we report it only as a warning.

\begin{table}[t]
\centering
\caption{Rejection rates of the global corner-averaged IPW CvM at \(\alpha=0.05\), hill
and flat surfaces, under the phenomenological canopy-return model. All cells are Monte
Carlo estimates from 100 replications, with \(B=99\) parametric-bootstrap draws under the
constant-mean-cover null \(p_{0}=0.5\), \(m_{\mathrm{ref}}=150\), grid \(350\times350\).
In panel (B), the flat constant and flat slope-dependent cells are null controls; the flat
hotspot cells are power against a non-constant alternative. \textbf{(A)} Hill, surface
design, three pulse densities. \textbf{(B)} \(n=5000\), both
surfaces and both sampling designs; ``plane'' draws pulse locations uniformly on the
parameter square and weights by \(\rho(X)/Z\) in addition to \(1/q_{c}(X)\), where
\(Z=\int_{D}\rho\). The column headed ``naive'' in panel (B) is the unweighted test,
reported only to show that its rejection rate tracks the design rather than the cover
signal.}
\label{tab:global-ipw-cvm}
\begin{tabular}{@{}rrrr@{}}
\toprule
\multicolumn{4}{@{}l}{\textbf{(A) Hill, surface design: power against pulse density}} \\
\addlinespace
$n$ & Size & Slope-dependent cover & Compact hotspot \\
\midrule
500  & 0.03 & 0.03 & 0.05 \\
2000 & 0.05 & 0.26 & 0.06 \\
5000 & 0.04 & 0.79 & 0.12 \\
\bottomrule
\end{tabular}\par\vspace{8pt}
\begin{tabular}{@{}llrrrr@{}}
\toprule
\multicolumn{6}{@{}l}{\textbf{(B) $n=5000$: surface against plane design}} \\
\addlinespace
Surface & Design & Size & Slope-dependent cover & Compact hotspot
        & Naive, slope-dependent \\
\midrule
Hill & surface & 0.04 & 0.79 & 0.12 & 1.00 \\
Hill & plane   & 0.04 & 0.85 & 0.13 & 0.16 \\
\addlinespace
Flat & surface & 0.02 & 0.03 & 0.09 & 0.06 \\
Flat & plane   & 0.04 & 0.06 & 0.10 & 0.06 \\
\bottomrule
\end{tabular}
\end{table}

\subsection{The plane-uniform arm}\label{sec:plane-results}

The surface-uniform design cancels $\rho$ by construction, so it exercises the detection weight
only. The plane-uniform arm samples pulses uniformly on $D$ and uses the explicit weights
$\rho(X_{i})/Z$ divided by $q_{c}(X_{i})$, which is closer to a real airborne design. At $n=5000$ on the hill it gives size $0.04$, power $0.85$
($0.765$--$0.914$) against slope-dependent cover, and $0.13$ against the hotspot; on the flat
surface it gives $0.04$, $0.06$ and $0.10$. The weighted estimator's bias stays below $0.003$ in
absolute value in this arm as well. The difference between $0.85$ and the surface-uniform $0.79$
is within Monte Carlo error at 100 replications (standard error of the difference about $0.05$),
and we do not interpret it.

Under plane-uniform sampling the naive planar statistic
has power $0.16$ on the hill slope alternative, not $1.00$. The $1.00$ in the surface-uniform
design reflects the detection field as much as the cover field, so the naive statistic is not
evidence about $p_{\mathrm{surf}}$ in either design.

\subsection{The local scan: a negative result}\label{sec:local}

\begin{figure}[t]
\centering
\includegraphics[width=\textwidth]{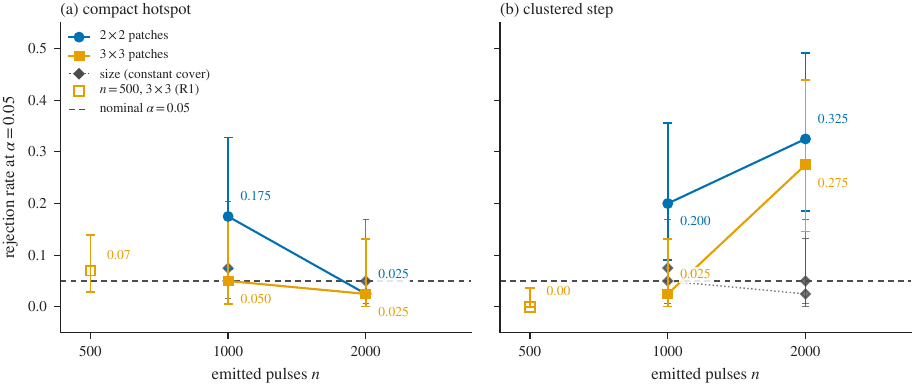}
\caption{Local max-composite scan on the hill under the phenomenological canopy-return
model: rejection rate at \(\alpha=0.05\) against pulse density for \(2\times2\) and
\(3\times3\) patch families, with the size of the scan (grey diamonds) for reference.
Panel (a): compact hotspot. Panel (b): clustered step. The open square at \(n=500\) is the
\(3\times3\) point from the separate study R1 (100 replications, \(m_{\mathrm{ref}}=60\));
all other points are from R3 (40 replications, \(B=50\), \(m_{\mathrm{ref}}=40\)). Bars are
exact Clopper--Pearson 95\,\% intervals, which at 40 replications span roughly a third of
the unit interval. Synthetic pilot, not field-validated; the caption's job is to stop the
eye from reading 0.175 and 0.025 as well measured.}
\label{fig:local-scan}
\end{figure}

Figure~\ref{fig:local-scan} plots the scan power, and Table~\ref{tab:local-scan} reports the
max-composite scan on the hill under realistic detection, with
$2\times2$ and $3\times3$ patches at $n=1000$ and $n=2000$ and 40 replications. In the order
$(n,k)=(1000,2),(1000,3),(2000,2),(2000,3)$, size is $0.075$, $0.050$, $0.025$ and $0.050$, and
hotspot power is $0.175$, $0.050$, $0.025$ and $0.025$. With the $n=500$, $3\times3$ run
included, hotspot power lies between $0.025$ and $0.175$ over $n=500$--$2000$. The largest value
is $7/40$, with a 95\% interval of $0.07$--$0.33$; the smallest is $1/40$, with
$0.001$--$0.13$. The study therefore shows an absence of usable power at these densities, not a
precise power of $0.025$ and not that local detection is impossible.

Two structural features of the design explain the result.
First, a compact hotspot is a mean shift. At $n=500$ the $z$-scan rejects in
$0.09$, the local composite CvM scan in $0.07$, and the global CvM in $0.03$ (R1, $m_{\mathrm{ref}}=60$; Table~5 uses $m_{\mathrm{ref}}=150$); the local CvM
scan adds little over the $z$-scan, and the global statistic is the wrong instrument for this
alternative. Second, a $2\times2$ patch dilutes a hotspot covering a tenth of
the area over a quarter of it, while a $3\times3$ patch contains too few canopy returns at the
implied return rate. Against an artificial clustered step the scan does reach $0.33$ at
$n=2000$ with $2\times2$ patches, so the negative result is specific to the compact mean shift.
Localisation rates are not reported: with $2\times2$ patches all four patches are labelled
central, which makes the recorded rate of $1.0$ degenerate. The requirements that a usable scan
would have to meet---a data-independent patch family, a maximum with a null law, and a
multiple-testing argument---are the standard ones
\citep{kulldorff1997spatial,tango2012flexible,dangelo2023local}; the present scan has the first
but neither of the other two.

\begin{table}[t]
\centering
\caption{Max-composite local scan on the hill under the phenomenological canopy-return
model. Rejection rates at \(\alpha=0.05\) with a parametric bootstrap. The \(n=500\) row
is from study R1 (100 replications, \(m_{\mathrm{ref}}=60\)); the \(n=1000\) and \(n=2000\)
rows are from study R3 (40 replications, \(B=50\), \(m_{\mathrm{ref}}=40\)). All cells are
Monte Carlo estimates. Localisation rates are omitted: with \(2\times2\) patches all four
patches intersect the central square, so the recorded localisation rate is degenerate.}
\label{tab:local-scan}
\begin{tabular}{@{}rrrrr@{}}
\toprule
$n$ & Patches & Size & Compact hotspot & Clustered step \\
\midrule
500  & $3\times3$ & 0.050 & 0.070 & 0.000 \\
\addlinespace
1000 & $2\times2$ & 0.075 & 0.175 & 0.200 \\
1000 & $3\times3$ & 0.050 & 0.050 & 0.025 \\
2000 & $2\times2$ & 0.025 & 0.025 & 0.325 \\
2000 & $3\times3$ & 0.050 & 0.025 & 0.275 \\
\bottomrule
\end{tabular}
\end{table}

\subsection{What the pilot does and does not establish}\label{sec:limits-results}

The evidence is a synthetic pilot, not a field validation. Six limitations bound it. (i) The
power study uses one surface geometry, one alternative strength ($\beta=0.25$), one hotspot size
($10\%$ of the area, $+0.2$) and one detection parameterisation. (ii) $q_{c}$ is known by construction:
the terrain-derived density proxy that drives it is independent of the realised binary canopy, which
makes the parametric bootstrap valid but simplifies real detection. (iii) The detection model is
phenomenological, not a radiative-transfer model. (iv) The bootstrap draws independent canopy
returns from $\mathrm{Bernoulli}(p_{0}q_{c})$ with $p_{0}=0.5$, so the calibration is model-based
and the null is simple, not composite; no design-based null law is available. (v) The reference
CDF is a Monte Carlo average over $m_{\mathrm{ref}}=150$ points redrawn at every replication, so
the computed statistic approximates the integral it defines. (vi) $5000$ pulses over the unit
square is a scaling statement about the simulated domain, not a survey design. Real LiDAR and
field-plot validation remain open, and no number in this section should be read as a property of
an operational scanner.

\section{Discussion}\label{sec:discussion}

\subsection{Two neighbouring quantities, one estimand}

Standard canopy cover is defined by vertical projection: a canopy return at $(u,v)$ is
credited with the map-plane element $du\,dv$ beneath it, and the reported fraction is the share
of that planar area under canopy — a quantity that \citet{korhonen2011airborne} distinguish from
angular canopy closure and leaf area index. Our estimand keeps the canopy indicator and replaces
the denominator with surface area, weighting each point by $\rho\,du\,dv$. The two coincide
unless cover co-varies with $\rho$: the identity
$p_{\mathrm{surf}}=p_{\mathrm{planar}}+\mathrm{Cov}_{\mathrm{planar}}(C,\rho)/E_{\mathrm{planar}}[\rho]$
makes this exact. The constant slope shows why steepness is not the mechanism: ratio $1.118$,
gap zero. Slope-dependent cover on the hill (ratio $1.248$) gives $+0.0276$. Rumple is a third
object, the outer canopy surface area over ground area \citep{kane2010comparisons}: a roughness
of the canopy envelope with no canopy–ground partition, unbounded above and hence not a proportion. No rumple value
was computed here, so the comparison is conceptual and future work; \citet{panek2026novel}
estimate the area itself, not the covered fraction.

\subsection{The detection model is the load-bearing assumption}

Since $R_i=C(X_i)\mathbf{1}\{\text{return recorded}\}$,
$E[R_i\mid X_i]=p(X_i)q_c(X_i)$, so the naive fraction estimates $E_\mu[p\,q_c]$; under constant
cover this is $p_0E_\mu[q_c]$ ($0.17$ flat, $0.20$ hill), predicting the $-0.40$ bias before any
fitting. With $q_c$ known, the weighted mean is the Horvitz–Thompson mean over emitted pulses
\citep{horvitz1952generalization} and the factor cancels. In the field $q_c$ must be modelled
from scan geometry, a canopy height model and a clumping factor: scan angle beyond about
$12^\circ$ \citep{ma2017comparison}, non-returning pulses \citep{liu2019improving},
clumping-driven detectability \citep{xu2026quantifying}. Replacing the phenomenological model would mean a gap-probability or waveform model of the kind developed by \citet{ni2001modeling}, in which return probability follows from canopy architecture rather than being imposed. The implied mean canopy detection probability of $0.17$--$0.20$, conditional on canopy being present, corresponds to roughly one canopy return in six to one in five pulses emitted over canopy, or about one in ten emitted pulses, a low-return regime; the bias scales as $p_0(1-\mathbb{E}_\mu[q_c])$ and would shrink in a higher-return survey. A \(3\times2\) sweep over \(q_0\in\{0.6,0.9,1.0\}\) and \(\eta_c\in\{0.7,0.9\}\) at \(n=5000\) with 200 replications varied the naive bias from \(-0.448\) to \(-0.388\) while the IPW bias stayed within \(0.004\) in absolute value; the bias tracked \(p_0(1-\mathbb{E}_\mu[q_c])\). The sweep is recorded in \texttt{detection\_sensitivity.json}.

The simulation sidesteps this: $q_c$ is generated from a terrain-derived density proxy
independent of the realised binary canopy. That independence makes $q_c$ known and the bootstrap
correctly specified, and it is the main simplification: real detectability depends on the
foliage that defines $C$, making $q_c$ endogenous to the estimand. Field work must break the
circle with external information, and misspecification is not second order: a constant $q$
raised the flat-surface size from $0.07$ to $0.10$ in the composite-scan pilot.

\subsection{What the global test does and does not detect}

The statistic integrates a squared $L^2$ distance over the domain, so it suits broad departures
and not compact ones. Against slope-dependent cover with $\beta=0.25$ on the hill at 5000
pulses, power is $0.79$ (surface-uniform) and $0.85$ (plane-uniform); against a 10\% hotspot
it is $0.12$ and $0.13$; size on the hill was $0.03$–$0.05$ across 500–5000 pulses. It
answers whether mean cover is constant under the independent-return model, not where it is
anomalous. Calibration is model-based: under constant mean cover the weighted CDF is centred
correctly under any correlation, but correlation changes the variance, so label permutation has
size $1.00$ under a thresholded field (correlation length $\ell=0.1$), whereas a shift or field bootstrap of the
kind reviewed by \citet{mrkvicka2020revisiting} restores size to $0.04$–$0.06$ in the
ideal-sampling study. The detection study instead uses a parametric bootstrap with independent
returns and $p_0=0.5$. The two are never combined, so there is no design-based null law and
the $p$-values are calibrated under a model — the weakest joint in the inference.

The local scan is reported because the global statistic could otherwise be read as a hotspot
detector. Maximum-composite hotspot power was $0.025$–$0.175$ at $n=1000$–$2000$, with size
$0.025$–$0.075$; the largest is 7 of 40 ($0.07$–$0.33$), so the study shows the absence of usable
power, not a precise power. A compact hotspot is a mean shift: at $n=500$ the rejection rates
are $0.09$ for the $z$-scan, $0.07$ for the local composite CvM scan and $0.03$ for the global
CvM; the local CvM scan adds little over the $z$-scan, and the global statistic is the wrong
instrument for this alternative. The $2\times2$ localization rate of $1.0$ is degenerate, and
no multiple-testing argument covers the maximum. The local test of
\citet{dangelo2023local} is the comparator, and the requirements of the scan literature
\citep{kulldorff1997spatial,tango2012flexible} are unmet.

\subsection{The Lean core is a specification, not a validation}

The core checks the objects, not the inference: the area density, finiteness and normalisation
of the surface measure, the flat reduction to the project's corner-averaged statistic, the
weighted reductions, the $1/q_{\min}$ and unit-interval bounds, and whole-patch recovery for the
local scan \citep{demoura2021lean4,mathlib2020lean}. It contains no null law, unbiasedness
theorem, consistency result or critical value. Its value is that the object cannot drift: a later
treatment that silently changed the area element, corner definition or weight would no longer
match the checked statements. It is a machine-checked specification and a reproducibility
device, not a statistical validation, and no general non-flat reduction to the cube statistic is
claimed.

\subsection{Field programme and threats to validity}

Validation should test the estimand, not demonstrate the method. It needs an ALS or GEDI tile
with scan geometry and return classification, a canopy height model resolving the terrain
surface, and field plots with cover, slope and clumping; NEON supplies the ALS and plot
layers, and co-registered ALS is the reference against which GEDI cover is checked
\citep{li2024tree,li2024vertical}. The analysis has four steps: estimate $q_c$ from scan
angle, range and a clumping model, validating it against observed return rates rather than
against cover; compute $p_{\mathrm{surf}}$, planar cover and rumple on the same plots; report the
gap with its
components, since a large area ratio with near-zero covariance is the case where the estimand
buys nothing; and test constant mean cover both with the parametric bootstrap and a
correlation-robust scheme.

The estimand still credits canopy vertically: the terrain point and the canopy indicator share the same $(u,v)$, and the surface-area denominator replaces the planar one without changing the projection direction. A fully oblique or overhang-aware definition would require a three-dimensional canopy representation; the present quantity is a terrain-area-weighted vertical projection, not a view-angle-dependent canopy closure.

Four threats bound the evidence. Scope: one hill, one $\beta$, one hotspot radius, one detection
model, with power recorded only along pulse density ($0.03$, $0.26$, $0.79$ at 500, 2000, 5000),
so each figure is a single cell of a larger design. Calibration: no design-based null law,
$p_0$ fixed at $0.5$, independent returns, correlation and detection never combined. Detection:
phenomenological, with known $q_c$ and an auxiliary field independent of the realised canopy.
Data: none are real; the $-0.40$ magnitude is a property of the simulated regime
($E_\mu[q_c]\approx0.17$–$0.20$), not an estimate for a scanner or forest type. The mechanism
transfers; the magnitude does not.

\section{Conclusion}\label{sec:conclusion}

Canopy cover on irregular terrain is a fraction of the land surface, not of the map plane. We
defined that fraction for a graph surface and showed that it differs from vertical-projection
cover by the planimetric covariance of cover with area density, divided by mean area density.
The identity says when the choice matters: a constant slope with an area ratio of $1.118$ leaves
the estimand unchanged, whereas slope-dependent cover on the hill moves it by $+0.0276$.

LiDAR records a canopy return only when detection succeeds, so the naive return fraction is
biased by the detection regime; in the synthetic study that bias was about $-0.40$. When the
canopy detection probability is known, the inverse-probability-weighted estimator removed the
first-order bias: at 5000 pulses the bias was at most $0.003$ in absolute value; in the
pulse-density study the largest absolute bias at 500 pulses was $0.021$, in the flat
slope-dependent null cell where the cover map is constant.

The weighted corner-averaged Cramér–von Mises test addresses a different question: whether
mean surface-area cover is constant under independent returns. At 5000 pulses it held size
between $0.02$ and $0.06$ across the simulated null cells and reached power $0.79$ under a
surface-uniform design and $0.85$ under a plane-uniform design against broad slope-dependent
cover; against a compact hotspot it reached $0.09$–$0.13$.

The surface-uniform definitions, their finite-sample bounds and their exact reductions are machine-checked in
Lean 4. The null law, the detection model and the field validation remain open, and every
estimate reported here is synthetic.

\backmatter

\section*{Acknowledgments}

The author thanks Dr. Welfredo Patungan for consultations on the Cram\'er--von Mises criterion, and the maintainers of Lean~4 and Mathlib.

\section*{Statements and Declarations}

\begin{itemize}
\item \textbf{Funding:} This research received no external funding. It was supported by the author's personal funds.
\item \textbf{Competing interests:} The author declares no competing interests.
\item \textbf{Data availability:} The synthetic study and the Lean formal core are available from the author on reasonable request. The development repository is \url{https://github.com/mrphilo420/Project-SteinGate}; it is private until publication and contains the simulation scripts and JSON summaries in \texttt{numerics\_data/surface\_cvm/}, the Lean modules in \texttt{lean/SteinGate/CvM/Surface*.lean}, and the axiom audit in \texttt{verify/AxiomAudit.lean} and \texttt{scripts/verify.sh}. No real LiDAR or field-plot data are used in this version; the real-data validation plan is described in the Discussion.
\item \textbf{Author contribution:} Marco Mandap: Conceptualization; Methodology; Software; Formal analysis; Investigation; Data curation; Writing --- original draft; Writing --- review and editing; Visualization; Project administration; Funding acquisition.
\item \textbf{Use of generative AI:} During the preparation of this work the author used AI-assisted coding and drafting tools for code generation, Lean~4 proof engineering, bibliographic checking, and manuscript editing. After using these tools, the author reviewed and edited the content as needed and takes full responsibility for the content of the publication.
\item \textbf{Ethics approval:} Not applicable. This study used synthetic data only and did not involve human participants, animals, or identifiable personal data.
\item \textbf{Consent to participate and publish:} Not applicable.
\end{itemize}
\begin{appendices}

\section{Replication details}
\label{app:replication}

The synthetic study is generated by the scripts in
\texttt{numerics\_data/surface\_cvm/}. The Lean audit is run with

\begin{verbatim}
lake build SteinGate.CvM.SurfaceIPWCvM
scripts/verify.sh
\end{verbatim}

The headline theorems are registered in \texttt{verify/AxiomAudit.lean}; each prints only
\texttt{[propext, Classical.choice, Quot.sound]}. The live output of
\texttt{scripts/verify.sh} is authoritative. The audit reported here was run on
3 November 2026.

The ideal-sampling pilot uses seed 20261102. The detection and pulse-density studies use
seed 20261103. The principal summaries are
\texttt{results.json}, \texttt{realistic\_detection\_results.json},
\texttt{global\_pulse\_density.json}, \texttt{multiscale\_rescue\_results.json}, and
\texttt{detection\_sensitivity.json}; the tables and figure captions name the exact JSON
paths used for each number.

\end{appendices}

\bibliography{sn-bibliography}


\begin{thebibliography}{40}
\ifx \bisbn   \undefined \def \bisbn  #1{ISBN #1}\fi
\ifx \binits  \undefined \def \binits#1{#1}\fi
\ifx \bauthor  \undefined \def \bauthor#1{#1}\fi
\ifx \batitle  \undefined \def \batitle#1{#1}\fi
\ifx \bjtitle  \undefined \def \bjtitle#1{#1}\fi
\ifx \bvolume  \undefined \def \bvolume#1{\textbf{#1}}\fi
\ifx \byear  \undefined \def \byear#1{#1}\fi
\ifx \bissue  \undefined \def \bissue#1{#1}\fi
\ifx \bfpage  \undefined \def \bfpage#1{#1}\fi
\ifx \blpage  \undefined \def \blpage #1{#1}\fi
\ifx \burl  \undefined \def \burl#1{\textsf{#1}}\fi
\ifx \doiurl  \undefined \def \doiurl#1{\url{https://doi.org/#1}}\fi
\ifx \betal  \undefined \def \betal{\textit{et al.}}\fi
\ifx \binstitute  \undefined \def \binstitute#1{#1}\fi
\ifx \binstitutionaled  \undefined \def \binstitutionaled#1{#1}\fi
\ifx \bctitle  \undefined \def \bctitle#1{#1}\fi
\ifx \beditor  \undefined \def \beditor#1{#1}\fi
\ifx \bpublisher  \undefined \def \bpublisher#1{#1}\fi
\ifx \bbtitle  \undefined \def \bbtitle#1{#1}\fi
\ifx \bedition  \undefined \def \bedition#1{#1}\fi
\ifx \bseriesno  \undefined \def \bseriesno#1{#1}\fi
\ifx \blocation  \undefined \def \blocation#1{#1}\fi
\ifx \bsertitle  \undefined \def \bsertitle#1{#1}\fi
\ifx \bsnm \undefined \def \bsnm#1{#1}\fi
\ifx \bsuffix \undefined \def \bsuffix#1{#1}\fi
\ifx \bparticle \undefined \def \bparticle#1{#1}\fi
\ifx \barticle \undefined \def \barticle#1{#1}\fi
\bibcommenthead
\ifx \bconfdate \undefined \def \bconfdate #1{#1}\fi
\ifx \botherref \undefined \def \botherref #1{#1}\fi
\ifx \url \undefined \def \url#1{\textsf{#1}}\fi
\ifx \bchapter \undefined \def \bchapter#1{#1}\fi
\ifx \bbook \undefined \def \bbook#1{#1}\fi
\ifx \bcomment \undefined \def \bcomment#1{#1}\fi
\ifx \oauthor \undefined \def \oauthor#1{#1}\fi
\ifx \citeauthoryear \undefined \def \citeauthoryear#1{#1}\fi
\ifx \endbibitem  \undefined \def \endbibitem {}\fi
\ifx \bconflocation  \undefined \def \bconflocation#1{#1}\fi
\ifx \arxivurl  \undefined \def \arxivurl#1{\textsf{#1}}\fi
\csname PreBibitemsHook\endcsname

\bibitem[\protect\citeauthoryear{Jennings et~al.}{1999}]{jennings1999canopy}
\begin{barticle}
\bauthor{\bsnm{Jennings}, \binits{S.B.}},
\bauthor{\bsnm{Brown}, \binits{N.D.}},
\bauthor{\bsnm{Sheil}, \binits{D.}}:
\batitle{Assessing forest canopies and understorey illumination: Canopy
  closure, canopy cover and other measures}.
\bjtitle{Forestry}
\bvolume{72}(\bissue{1}),
\bfpage{59}--\blpage{74}
(\byear{1999})
\doiurl{10.1093/forestry/72.1.59}
\end{barticle}
\endbibitem

\bibitem[\protect\citeauthoryear{Korhonen et~al.}{2011}]{korhonen2011airborne}
\begin{barticle}
\bauthor{\bsnm{Korhonen}, \binits{L.}},
\bauthor{\bsnm{Korpela}, \binits{I.}},
\bauthor{\bsnm{Heiskanen}, \binits{J.}},
\bauthor{\bsnm{Maltamo}, \binits{M.}}:
\batitle{Airborne discrete-return {LIDAR} data in the estimation of vertical
  canopy cover, angular canopy closure and leaf area index}.
\bjtitle{Remote Sensing of Environment}
\bvolume{115}(\bissue{4}),
\bfpage{1065}--\blpage{1080}
(\byear{2011})
\doiurl{10.1016/j.rse.2010.12.011}
\end{barticle}
\endbibitem

\bibitem[\protect\citeauthoryear{Li et~al.}{2024}]{li2024vertical}
\begin{barticle}
\bauthor{\bsnm{Li}, \binits{Y.}},
\bauthor{\bsnm{Fang}, \binits{H.}},
\bauthor{\bsnm{Wang}, \binits{Y.}},
\bauthor{\bsnm{Li}, \binits{S.}},
\bauthor{\bsnm{Ma}, \binits{T.}},
\bauthor{\bsnm{Wu}, \binits{Y.}}:
\batitle{Validation of the vertical canopy cover profile products derived from
  {GEDI} over selected forest sites}.
\bjtitle{Science of Remote Sensing}
\bvolume{10},
\bfpage{100158}
(\byear{2024})
\doiurl{10.1016/j.srs.2024.100158}
\end{barticle}
\endbibitem

\bibitem[\protect\citeauthoryear{Panek-Chwastyk and
  Chwastyk}{2026}]{panek2026novel}
\begin{barticle}
\bauthor{\bsnm{Panek-Chwastyk}, \binits{E.}},
\bauthor{\bsnm{Chwastyk}, \binits{M.}}:
\batitle{A novel geometric approach to surface area estimation in forested
  mountain terrain using multi-source elevation and canopy height data}.
\bjtitle{Scientific Reports}
(\byear{2026})
\doiurl{10.1038/s41598-026-65204-1}
\end{barticle}
\endbibitem

\bibitem[\protect\citeauthoryear{Ma et~al.}{2017}]{ma2017comparison}
\begin{barticle}
\bauthor{\bsnm{Ma}, \binits{Q.}},
\bauthor{\bsnm{Su}, \binits{Y.}},
\bauthor{\bsnm{Guo}, \binits{Q.}}:
\batitle{Comparison of canopy cover estimations from airborne {LiDAR}, aerial
  imagery, and satellite imagery}.
\bjtitle{IEEE Journal of Selected Topics in Applied Earth Observations and
  Remote Sensing}
\bvolume{10}(\bissue{9}),
\bfpage{4225}--\blpage{4236}
(\byear{2017})
\doiurl{10.1109/JSTARS.2017.2711482}
\end{barticle}
\endbibitem

\bibitem[\protect\citeauthoryear{Dayal et~al.}{2022}]{dayal2022investigation}
\begin{barticle}
\bauthor{\bsnm{Dayal}, \binits{K.R.}},
\bauthor{\bsnm{Durrieu}, \binits{S.}},
\bauthor{\bsnm{Lahssini}, \binits{K.}},
\bauthor{\bsnm{Alleaume}, \binits{S.}},
\bauthor{\bsnm{Bouvier}, \binits{M.}},
\bauthor{\bsnm{Monnet}, \binits{J.-M.}}:
\batitle{An investigation into lidar scan angle impacts on stand attribute
  predictions in different forest environments}.
\bjtitle{ISPRS Journal of Photogrammetry and Remote Sensing}
\bvolume{193},
\bfpage{314}--\blpage{338}
(\byear{2022})
\doiurl{10.1016/j.isprsjprs.2022.08.013}
\end{barticle}
\endbibitem

\bibitem[\protect\citeauthoryear{Liu et~al.}{2020}]{liu2019improving}
\begin{barticle}
\bauthor{\bsnm{Liu}, \binits{Q.}},
\bauthor{\bsnm{Fu}, \binits{L.}},
\bauthor{\bsnm{Wang}, \binits{G.}},
\bauthor{\bsnm{Li}, \binits{S.}},
\bauthor{\bsnm{Li}, \binits{Z.}},
\bauthor{\bsnm{Chen}, \binits{E.}}:
\batitle{Improving estimation of forest canopy cover by introducing loss ratio
  of laser pulses using airborne {LiDAR}}.
\bjtitle{IEEE Transactions on Geoscience and Remote Sensing}
\bvolume{58}(\bissue{1}),
\bfpage{567}--\blpage{585}
(\byear{2020})
\doiurl{10.1109/TGRS.2019.2938017}
\end{barticle}
\endbibitem

\bibitem[\protect\citeauthoryear{Xu et~al.}{2026}]{xu2026quantifying}
\begin{botherref}
\oauthor{\bsnm{Xu}, \binits{X.}},
\oauthor{\bsnm{Liu}, \binits{X.}},
\oauthor{\bsnm{Guo}, \binits{J.}},
\oauthor{\bsnm{Wang}, \binits{X.}}:
Quantifying canopy-clumping effects on {LiDAR} detectability in heterogeneous
  forests using an effective detection factor.
International Journal of Remote Sensing,
1--39
(2026)
\doiurl{10.1080/01431161.2026.2734355}
\end{botherref}
\endbibitem

\bibitem[\protect\citeauthoryear{Park et~al.}{2014}]{park2014application}
\begin{barticle}
\bauthor{\bsnm{Park}, \binits{T.}},
\bauthor{\bsnm{Kennedy}, \binits{R.E.}},
\bauthor{\bsnm{Choi}, \binits{S.}},
\bauthor{\bsnm{Wu}, \binits{J.}},
\bauthor{\bsnm{Lefsky}, \binits{M.A.}},
\bauthor{\bsnm{Bi}, \binits{J.}}:
\batitle{Application of physically-based slope correction for maximum forest
  canopy height estimation using waveform lidar across different footprint
  sizes and locations: Tests on {LVIS} and {GLAS}}.
\bjtitle{Remote Sensing}
\bvolume{6}(\bissue{7}),
\bfpage{6566}--\blpage{6586}
(\byear{2014})
\doiurl{10.3390/rs6076566}
\end{barticle}
\endbibitem

\bibitem[\protect\citeauthoryear{Li et~al.}{2023}]{li2023topography}
\begin{barticle}
\bauthor{\bsnm{Li}, \binits{Y.}},
\bauthor{\bsnm{Zeng}, \binits{H.}},
\bauthor{\bsnm{Xiong}, \binits{J.}},
\bauthor{\bsnm{Miao}, \binits{G.}}:
\batitle{Influence of topography on {UAV} {LiDAR}-based {LAI} estimation in
  subtropical mountainous secondary broadleaf forests}.
\bjtitle{Forests}
\bvolume{15}(\bissue{1}),
\bfpage{17}
(\byear{2023})
\doiurl{10.3390/f15010017}
\end{barticle}
\endbibitem

\bibitem[\protect\citeauthoryear{Hou et~al.}{2026}]{hou2026slope}
\begin{bchapter}
\bauthor{\bsnm{Hou}, \binits{J.}},
\bauthor{\bsnm{Liao}, \binits{Z.}},
\bauthor{\bsnm{Wang}, \binits{J.}},
\bauthor{\bsnm{Wu}, \binits{L.}},
\bauthor{\bsnm{Chen}, \binits{Y.}},
\bauthor{\bsnm{Zhang}, \binits{Y.}}:
\bctitle{A slope-adaptive correction method for ground return to improve {GEDI}
  canopy cover estimation}.
In: \bbtitle{{IGARSS} 2026},
pp. \bfpage{3502}--\blpage{3505}
(\byear{2026}).
\doiurl{10.1109/IGARSS60577.2026.11730818}
\end{bchapter}
\endbibitem

\bibitem[\protect\citeauthoryear{Schneider
  et~al.}{2019}]{schneider2019quantifying}
\begin{barticle}
\bauthor{\bsnm{Schneider}, \binits{F.}},
\bauthor{\bsnm{K{\"u}kenbrink}, \binits{D.}},
\bauthor{\bsnm{Schaepman}, \binits{M.E.}},
\bauthor{\bsnm{Schimel}, \binits{D.S.}},
\bauthor{\bsnm{Morsdorf}, \binits{F.}}:
\batitle{Quantifying {3D} structure and occlusion in dense tropical and
  temperate forests using close-range {LiDAR}}.
\bjtitle{Agricultural and Forest Meteorology}
\bvolume{268},
\bfpage{249}--\blpage{257}
(\byear{2019})
\doiurl{10.1016/j.agrformet.2019.01.033}
\end{barticle}
\endbibitem

\bibitem[\protect\citeauthoryear{Donager
  et~al.}{2021}]{donager2021adjudicating}
\begin{barticle}
\bauthor{\bsnm{Donager}, \binits{J.J.}},
\bauthor{\bsnm{S{\'a}nchez~Meador}, \binits{A.J.}},
\bauthor{\bsnm{Blackburn}, \binits{R.C.}}:
\batitle{Adjudicating perspectives on forest structure: How do airborne,
  terrestrial, and mobile lidar-derived estimates compare?}
\bjtitle{Remote Sensing}
\bvolume{13}(\bissue{12}),
\bfpage{2297}
(\byear{2021})
\doiurl{10.3390/rs13122297}
\end{barticle}
\endbibitem

\bibitem[\protect\citeauthoryear{{\"O}zt{\"u}rk}{1997}]{ozturk1997generalised}
\begin{barticle}
\bauthor{\bsnm{{\"O}zt{\"u}rk}, \binits{{\"O}.}}:
\batitle{Generalised weighted {C}ram{\'e}r--von {M}ises distance estimators}.
\bjtitle{Biometrika}
\bvolume{84}(\bissue{2}),
\bfpage{283}--\blpage{294}
(\byear{1997})
\doiurl{10.1093/biomet/84.2.283}
\end{barticle}
\endbibitem

\bibitem[\protect\citeauthoryear{Martynov}{2011}]{martynov2011weighted}
\begin{barticle}
\bauthor{\bsnm{Martynov}, \binits{G.V.}}:
\batitle{Weighted {C}ram{\'e}r--von {M}ises test with estimated parameters}.
\bjtitle{Communications in Statistics -- Theory and Methods}
\bvolume{40}(\bissue{19--20}),
\bfpage{3569}--\blpage{3586}
(\byear{2011})
\doiurl{10.1080/03610926.2011.581179}
\end{barticle}
\endbibitem

\bibitem[\protect\citeauthoryear{Jupp}{2005}]{jupp2005sobolev}
\begin{barticle}
\bauthor{\bsnm{Jupp}, \binits{P.E.}}:
\batitle{Sobolev tests of goodness of fit of distributions on compact
  {R}iemannian manifolds}.
\bjtitle{The Annals of Statistics}
\bvolume{33}(\bissue{6}),
\bfpage{2957}--\blpage{2966}
(\byear{2005})
\doiurl{10.1214/009053605000000697}
\end{barticle}
\endbibitem

\bibitem[\protect\citeauthoryear{Jupp and Kume}{2020}]{jupp2020measures}
\begin{barticle}
\bauthor{\bsnm{Jupp}, \binits{P.E.}},
\bauthor{\bsnm{Kume}, \binits{A.}}:
\batitle{Measures of goodness of fit obtained by almost-canonical
  transformations on {R}iemannian manifolds}.
\bjtitle{Journal of Multivariate Analysis}
\bvolume{176},
\bfpage{104579}
(\byear{2020})
\doiurl{10.1016/j.jmva.2019.104579}
\end{barticle}
\endbibitem

\bibitem[\protect\citeauthoryear{Garc{\'i}a-Portugu{\'e}s
  et~al.}{2021}]{garcia2021cvm}
\begin{bchapter}
\bauthor{\bsnm{Garc{\'i}a-Portugu{\'e}s}, \binits{E.}},
\bauthor{\bsnm{Navarro-Esteban}, \binits{P.}},
\bauthor{\bsnm{Cuesta-Albertos}, \binits{J.A.}}:
\bctitle{A {C}ram{\'e}r--von {M}ises test of uniformity on the hypersphere}.
In: \bbtitle{Studies in Classification, Data Analysis, and Knowledge
  Organization},
pp. \bfpage{107}--\blpage{116}
(\byear{2021}).
\doiurl{10.1007/978-3-030-69944-4_12}
\end{bchapter}
\endbibitem

\bibitem[\protect\citeauthoryear{Zimmerman}{1993}]{zimmerman1993bivariate}
\begin{barticle}
\bauthor{\bsnm{Zimmerman}, \binits{D.L.}}:
\batitle{A bivariate {C}ram{\'e}r--von {M}ises type of test for spatial
  randomness}.
\bjtitle{Journal of the Royal Statistical Society: Series C (Applied
  Statistics)}
\bvolume{42}(\bissue{1}),
\bfpage{43}
(\byear{1993})
\doiurl{10.2307/2347408}
\end{barticle}
\endbibitem

\bibitem[\protect\citeauthoryear{Schoenberg}{2004}]{schoenberg2004testing}
\begin{barticle}
\bauthor{\bsnm{Schoenberg}, \binits{F.P.}}:
\batitle{Testing separability in spatial-temporal marked point processes}.
\bjtitle{Biometrics}
\bvolume{60}(\bissue{2}),
\bfpage{471}--\blpage{481}
(\byear{2004})
\doiurl{10.1111/j.0006-341X.2004.00192.x}
\end{barticle}
\endbibitem

\bibitem[\protect\citeauthoryear{Grabarnik et~al.}{2011}]{grabarnik2011correct}
\begin{barticle}
\bauthor{\bsnm{Grabarnik}, \binits{P.}},
\bauthor{\bsnm{Myllym{\"a}ki}, \binits{M.}},
\bauthor{\bsnm{Stoyan}, \binits{D.}}:
\batitle{Correct testing of mark independence for marked point patterns}.
\bjtitle{Ecological Modelling}
\bvolume{222}(\bissue{23--24}),
\bfpage{3888}--\blpage{3894}
(\byear{2011})
\doiurl{10.1016/j.ecolmodel.2011.10.005}
\end{barticle}
\endbibitem

\bibitem[\protect\citeauthoryear{Myllym{\"a}ki
  et~al.}{2015}]{myllymaki2015deviation}
\begin{barticle}
\bauthor{\bsnm{Myllym{\"a}ki}, \binits{M.}},
\bauthor{\bsnm{Grabarnik}, \binits{P.}},
\bauthor{\bsnm{Seijo}, \binits{H.}},
\bauthor{\bsnm{Stoyan}, \binits{D.}}:
\batitle{Deviation test construction and power comparison for marked spatial
  point patterns}.
\bjtitle{Spatial Statistics}
\bvolume{11},
\bfpage{19}--\blpage{34}
(\byear{2015})
\doiurl{10.1016/j.spasta.2014.11.004}
\end{barticle}
\endbibitem

\bibitem[\protect\citeauthoryear{Mrkvi{\v{c}}ka
  et~al.}{2012}]{mrkvicka2012goodness}
\begin{barticle}
\bauthor{\bsnm{Mrkvi{\v{c}}ka}, \binits{T.}},
\bauthor{\bsnm{Soubeyrand}, \binits{S.}},
\bauthor{\bsnm{Chad{\oe}uf}, \binits{J.}}:
\batitle{Goodness-of-fit test of the mark distribution in a point process with
  non-stationary marks}.
\bjtitle{Statistics and Computing}
\bvolume{22}(\bissue{4}),
\bfpage{931}--\blpage{943}
(\byear{2012})
\doiurl{10.1007/s11222-011-9263-y}
\end{barticle}
\endbibitem

\bibitem[\protect\citeauthoryear{Heinrich
  et~al.}{2014}]{heinrich2014asymptotic}
\begin{barticle}
\bauthor{\bsnm{Heinrich}, \binits{L.}},
\bauthor{\bsnm{L{\"u}ck}, \binits{S.}},
\bauthor{\bsnm{Schmidt}, \binits{V.}}:
\batitle{Asymptotic goodness-of-fit tests for the {Palm} mark distribution of
  stationary point processes with correlated marks}.
\bjtitle{Bernoulli}
\bvolume{20}(\bissue{4}),
\bfpage{1673}--\blpage{1697}
(\byear{2014})
\doiurl{10.3150/13-BEJ523}
\end{barticle}
\endbibitem

\bibitem[\protect\citeauthoryear{Horvitz and
  Thompson}{1952}]{horvitz1952generalization}
\begin{barticle}
\bauthor{\bsnm{Horvitz}, \binits{D.G.}},
\bauthor{\bsnm{Thompson}, \binits{D.J.}}:
\batitle{A generalization of sampling without replacement from a finite
  universe}.
\bjtitle{Journal of the American Statistical Association}
\bvolume{47},
\bfpage{663}--\blpage{685}
(\byear{1952})
\doiurl{10.1080/01621459.1952.10483446}
\end{barticle}
\endbibitem

\bibitem[\protect\citeauthoryear{{de Moura} and
  Ullrich}{2021}]{demoura2021lean4}
\begin{bchapter}
\bauthor{\bsnm{{de Moura}}, \binits{L.}},
\bauthor{\bsnm{Ullrich}, \binits{S.}}:
\bctitle{The {Lean} 4 theorem prover and programming language}.
In: \bbtitle{Automated Deduction -- CADE 28}.
\bsertitle{Lecture Notes in Computer Science},
pp. \bfpage{625}--\blpage{635}
(\byear{2021}).
\doiurl{10.1007/978-3-030-79876-5_37}
\end{bchapter}
\endbibitem

\bibitem[\protect\citeauthoryear{{The Mathlib
  Community}}{2020}]{mathlib2020lean}
\begin{bchapter}
\bauthor{\bsnm{{The Mathlib Community}}}:
\bctitle{The {Lean} mathematical library}.
In: \bbtitle{Proceedings of the 9th ACM SIGPLAN International Conference on
  Certified Programs and Proofs},
pp. \bfpage{367}--\blpage{381}
(\byear{2020}).
\doiurl{10.1145/3372885.3373824}
\end{bchapter}
\endbibitem

\bibitem[\protect\citeauthoryear{Jenness}{2004}]{jenness2004calculating}
\begin{barticle}
\bauthor{\bsnm{Jenness}, \binits{J.S.}}:
\batitle{Calculating landscape surface area from digital elevation models}.
\bjtitle{Wildlife Society Bulletin}
\bvolume{32}(\bissue{3}),
\bfpage{829}--\blpage{839}
(\byear{2004})
\doiurl{10.2193/0091-7648(2004)032[0829:CLSAFD]2.0.CO;2}
\end{barticle}
\endbibitem

\bibitem[\protect\citeauthoryear{Kane et~al.}{2010}]{kane2010comparisons}
\begin{barticle}
\bauthor{\bsnm{Kane}, \binits{V.R.}},
\bauthor{\bsnm{McGaughey}, \binits{R.J.}},
\bauthor{\bsnm{Bakker}, \binits{J.D.}},
\bauthor{\bsnm{Gersonde}, \binits{R.F.}},
\bauthor{\bsnm{Lutz}, \binits{J.A.}},
\bauthor{\bsnm{Franklin}, \binits{J.F.}}:
\batitle{Comparisons between field- and {LiDAR}-based measures of stand
  structural complexity}.
\bjtitle{Canadian Journal of Forest Research}
\bvolume{40}(\bissue{4}),
\bfpage{761}--\blpage{773}
(\byear{2010})
\doiurl{10.1139/X10-024}
\end{barticle}
\endbibitem

\bibitem[\protect\citeauthoryear{St{\aa}hl et~al.}{2016}]{stahl2016models}
\begin{barticle}
\bauthor{\bsnm{St{\aa}hl}, \binits{G.}},
\bauthor{\bsnm{Saarela}, \binits{S.}},
\bauthor{\bsnm{Schnell}, \binits{S.}},
\bauthor{\bsnm{Holm}, \binits{S.}},
\bauthor{\bsnm{Breidenbach}, \binits{J.}},
\bauthor{\bsnm{Healey}, \binits{S.P.}},
\bauthor{\bsnm{Patterson}, \binits{P.L.}},
\bauthor{\bsnm{Magnussen}, \binits{S.}},
\bauthor{\bsnm{N{\ae}sset}, \binits{E.}},
\bauthor{\bsnm{McRoberts}, \binits{R.E.}},
\bauthor{\bsnm{Gregoire}, \binits{T.G.}}:
\batitle{Use of models in large-area forest surveys: Comparing model-assisted,
  model-based and hybrid estimation}.
\bjtitle{Forest Ecosystems}
\bvolume{3}(\bissue{1}),
\bfpage{5}
(\byear{2016})
\doiurl{10.1186/s40663-016-0064-9}
\end{barticle}
\endbibitem

\bibitem[\protect\citeauthoryear{Hopkinson and
  Chasmer}{2009}]{hopkinson2009testing}
\begin{barticle}
\bauthor{\bsnm{Hopkinson}, \binits{C.}},
\bauthor{\bsnm{Chasmer}, \binits{L.}}:
\batitle{Testing {LiDAR} models of fractional cover across multiple forest
  ecozones}.
\bjtitle{Remote Sensing of Environment}
\bvolume{113}(\bissue{1}),
\bfpage{275}--\blpage{288}
(\byear{2009})
\doiurl{10.1016/j.rse.2008.09.012}
\end{barticle}
\endbibitem

\bibitem[\protect\citeauthoryear{Kang and Schafer}{2007}]{kang2007demystifying}
\begin{barticle}
\bauthor{\bsnm{Kang}, \binits{J.D.Y.}},
\bauthor{\bsnm{Schafer}, \binits{J.L.}}:
\batitle{Demystifying double robustness: A comparison of alternative strategies
  for estimating a population mean from incomplete data}.
\bjtitle{Statistical Science}
\bvolume{22}(\bissue{4}),
\bfpage{523}--\blpage{539}
(\byear{2007})
\doiurl{10.1214/07-STS227}
\end{barticle}
\endbibitem

\bibitem[\protect\citeauthoryear{Mrkvi{\v{c}}ka
  et~al.}{2020}]{mrkvicka2020revisiting}
\begin{barticle}
\bauthor{\bsnm{Mrkvi{\v{c}}ka}, \binits{T.}},
\bauthor{\bsnm{Dvo{\v{r}}{\'a}k}, \binits{J.}},
\bauthor{\bsnm{Gonz{\'a}lez}, \binits{J.A.}},
\bauthor{\bsnm{Mateu}, \binits{J.}}:
\batitle{Revisiting the random shift approach for testing in spatial
  statistics}.
\bjtitle{Spatial Statistics}
\bvolume{42},
\bfpage{100430}
(\byear{2020})
\doiurl{10.1016/j.spasta.2020.100430}
\end{barticle}
\endbibitem

\bibitem[\protect\citeauthoryear{Radersma and Sheldon}{2015}]{radersma2015new}
\begin{barticle}
\bauthor{\bsnm{Radersma}, \binits{R.}},
\bauthor{\bsnm{Sheldon}, \binits{B.C.}}:
\batitle{A new permutation technique to explore and control for spatial
  autocorrelation}.
\bjtitle{Methods in Ecology and Evolution}
\bvolume{6}(\bissue{9}),
\bfpage{1026}--\blpage{1033}
(\byear{2015})
\doiurl{10.1111/2041-210X.12390}
\end{barticle}
\endbibitem

\bibitem[\protect\citeauthoryear{D'Angelo et~al.}{2023}]{dangelo2023local}
\begin{barticle}
\bauthor{\bsnm{D'Angelo}, \binits{N.}},
\bauthor{\bsnm{Adelfio}, \binits{G.}},
\bauthor{\bsnm{Mateu}, \binits{J.}},
\bauthor{\bsnm{Cronie}, \binits{O.}}:
\batitle{Local inhomogeneous weighted summary statistics for marked point
  processes}.
\bjtitle{Journal of Computational and Graphical Statistics}
\bvolume{33}(\bissue{2}),
\bfpage{588}--\blpage{602}
(\byear{2023})
\doiurl{10.1080/10618600.2023.2206441}
\end{barticle}
\endbibitem

\bibitem[\protect\citeauthoryear{Avigad et~al.}{2017}]{avigad2017formally}
\begin{barticle}
\bauthor{\bsnm{Avigad}, \binits{J.}},
\bauthor{\bsnm{H{\"o}lzl}, \binits{J.}},
\bauthor{\bsnm{Serafin}, \binits{L.}}:
\batitle{A formally verified proof of the central limit theorem}.
\bjtitle{Journal of Automated Reasoning}
\bvolume{59}(\bissue{4}),
\bfpage{389}--\blpage{423}
(\byear{2017})
\doiurl{10.1007/s10817-017-9404-x}
\end{barticle}
\endbibitem

\bibitem[\protect\citeauthoryear{Kulldorff}{1997}]{kulldorff1997spatial}
\begin{barticle}
\bauthor{\bsnm{Kulldorff}, \binits{M.}}:
\batitle{A spatial scan statistic}.
\bjtitle{Communications in Statistics -- Theory and Methods}
\bvolume{26}(\bissue{6}),
\bfpage{1481}--\blpage{1496}
(\byear{1997})
\doiurl{10.1080/03610929708831995}
\end{barticle}
\endbibitem

\bibitem[\protect\citeauthoryear{Tango and Takahashi}{2012}]{tango2012flexible}
\begin{barticle}
\bauthor{\bsnm{Tango}, \binits{T.}},
\bauthor{\bsnm{Takahashi}, \binits{K.}}:
\batitle{A flexible spatial scan statistic with a restricted likelihood ratio
  for detecting disease clusters}.
\bjtitle{Statistics in Medicine}
\bvolume{31}(\bissue{30}),
\bfpage{4207}--\blpage{4218}
(\byear{2012})
\doiurl{10.1002/sim.5478}
\end{barticle}
\endbibitem

\bibitem[\protect\citeauthoryear{Ni-Meister et~al.}{2001}]{ni2001modeling}
\begin{barticle}
\bauthor{\bsnm{Ni-Meister}, \binits{W.}},
\bauthor{\bsnm{Jupp}, \binits{D.L.B.}},
\bauthor{\bsnm{Dubayah}, \binits{R.}}:
\batitle{Modeling lidar waveforms in heterogeneous and discrete canopies}.
\bjtitle{IEEE Transactions on Geoscience and Remote Sensing}
\bvolume{39}(\bissue{9}),
\bfpage{1943}--\blpage{1958}
(\byear{2001})
\doiurl{10.1109/36.951085}
\end{barticle}
\endbibitem

\bibitem[\protect\citeauthoryear{Li et~al.}{2024}]{li2024tree}
\begin{barticle}
\bauthor{\bsnm{Li}, \binits{X.}},
\bauthor{\bsnm{Li}, \binits{L.}},
\bauthor{\bsnm{Mu}, \binits{X.}}, \betal:
\batitle{Validating {GEDI} tree canopy cover product across forest types using
  co-registered aerial {LiDAR} data}.
\bjtitle{ISPRS Journal of Photogrammetry and Remote Sensing}
\bvolume{207},
\bfpage{326}--\blpage{337}
(\byear{2024})
\doiurl{10.1016/j.isprsjprs.2023.11.024}
\end{barticle}
\endbibitem

\end{thebibliography}

\end{document}